\documentclass[aoas,preprint]{imsart}

\usepackage[parfill]{parskip}
\usepackage{graphicx}
\usepackage{amssymb}
\usepackage{float}
\usepackage{amsfonts}
\usepackage{amsmath}
\usepackage{epstopdf}
\usepackage{amsthm}
\usepackage{tikz}
\usepackage{pgfplots}
\usepackage[english]{babel}
\usepackage{dsfont}
\usepackage{geometry}
\usepackage[mathscr]{euscript}
\usepackage{bbm}
\usepackage{hyperref}
\usepackage[ruled]{algorithm2e}
\usepackage{longtable}
\RequirePackage[authoryear]{natbib}

\startlocaldefs
\newtheorem{theorem}{Theorem}
\newtheorem{lemma}{Lemma}

\newcommand{\changes}[1]{{\color{black}#1}}

\newcommand{\ms}{\mbox{m s$^{-1}$}}

\endlocaldefs

\begin{document}

\begin{frontmatter}

\title{A Hermite-Gaussian Based Radial Velocity Estimation Method }
\runtitle{Hermite-Gaussian Estimation Method}

\author{\fnms{Parker} \snm{Holzer}\ead[label=e1]{parker.holzer@yale.edu}\thanksref{m1}}
\and
\author{\fnms{Jessi} \snm{Cisewski-Kehe}\ead[label=e2]{jessica.cisewski@yale.edu}\thanksref{m1}}
\and
\author{\fnms{Debra} \snm{Fischer}\ead[label=e3]{debra.fischer@yale.edu}\thanksref{m2}}
\and
\author{\fnms{Lily} \snm{Zhao}\ead[label=e4]{lily.zhao@yale.edu}\thanksref{m2}
}

\affiliation{Department of Statistics \& Data Science, Yale University\thanksmark{m1}}
\affiliation{Department of Astronomy, Yale University\thanksmark{m2}}
\address{\printead{e1,e2,e3,e4}}

\runauthor{P. Holzer et al.}

\begin{abstract}
\ As the first successful technique used to detect exoplanets orbiting distant stars, the Radial Velocity Method aims to detect a periodic Doppler shift in a star's spectrum.  We introduce a new, mathematically rigorous, approach \changes{to detect such a signal} that accounts for \changes{functional relationships of} neighboring wavelengths, minimizes the role of wavelength interpolation, \changes{accounts for heteroskedastic noise}, \changes{and} easily allows for statistical inference. Using Hermite-Gaussian functions, we show that the problem of detecting a Doppler shift in the spectrum can be reduced to linear regression in many settings. \changes{A simulation study demonstrates that the proposed method is able to accurately estimate an individual spectrum's radial velocity with precision below $0.3$ \ms. Furthermore, the new method outperforms the traditional Cross-Correlation Function approach by reducing the root mean squared error up to $15$ cm s$^{-1}$.} The proposed method is \changes{also} demonstrated on a new set of observations \changes{from the EXtreme PREcision  Spectrometer  (EXPRES)} for the star 51 Pegasi, and successfully recovers estimates that agree well with previous studies of this planetary system. Data and \changes{Python3} code associated with this work can be found at \href{https://github.com/parkerholzer/hgrv_method}{$ https://github.com/parkerholzer/hgrv\underline{ \ }method $}.  \changes{The method is also implemented in the open source R package \textit{rvmethod}}.
\end{abstract}

%\begin{keyword}[class=MSC]
%\kwd[Primary ]{}
%\kwd{}
%\kwd[; secondary ]{}
%\end{keyword}

%\begin{keyword}
%\kwd{}
%\kwd{}
%\end{keyword}

\end{frontmatter}

% AOS,AOAS: If there are supplements please fill:
%\begin{supplement}[id=suppA]
%  \sname{Supplement A}
%  \stitle{Title}
%  \slink[doi]{10.1214/00-AOASXXXXSUPP}
%  \sdatatype{.pdf}" 
%  \sdescription{Some text}
%\end{supplement}

\section{Introduction}
\begin{figure}[ht!]
\centering
\includegraphics[width=0.7\linewidth]{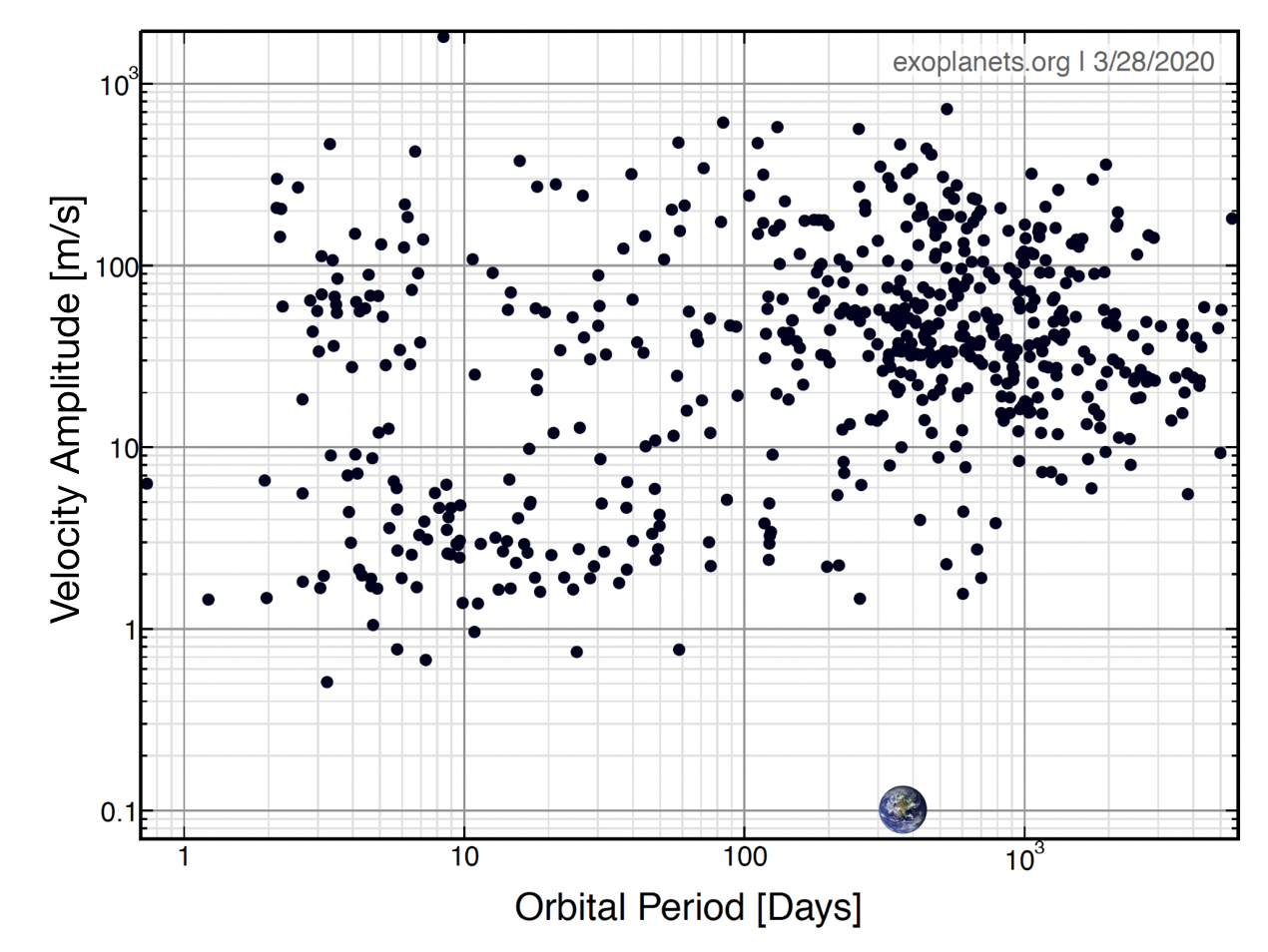}
\caption{Orbital period and stellar RV semi-amplitude for all exoplanets discovered with the RV Method. Data come from Exoplanets.org \citep{Han2014} \changes{on March 28, 2020 with a total of about $800$ exoplanets}. Note that with an orbital period of $365.25$ days and a semi-amplitude of approximately $0.1$ \ms, analogs of the Earth were not detectable.}
\label{fig:exoplans}
\end{figure}

\changes{The discovery of a planet orbiting the Sun-like star 51 Pegasi \citep{mayor95} launched a new subfield in astronomy: the detection and characterization of planets orbiting other stars, or exoplanets.
This discovery was made using the radial velocity (RV) method (also known as the Doppler technique). The RV method makes use of stellar spectra to derive the radial component of stellar velocity over time. Orbiting planets will tug the star around a common center of mass, producing a cyclical variation in the velocity of the target star with the same period as the orbiting planet. 

The data for the RV method are obtained with a spectrograph. The optical elements in the spectrograph disperse light from the star into component wavelengths, and focus the spectrum onto an electronic detector. The pixels in the detector sample the stellar spectrum. The continuous stellar spectrum is imprinted with thousands of narrow absorption lines that form when atoms and molecules in the outer atmosphere (hereafter referred to as the photosphere) of the star absorb specific wavelengths of light, corresponding to the quantum mechanical energy level differences in the absorbing atoms. As the star moves toward us or away from us, the velocity component that is projected along our line of site, i.e., the radial velocity, produces a wavelength rescaling in the spectrum that is described by the Doppler equation. 

All stars orbit the galaxy and will exhibit a nearly constant radial velocity relative to the Sun. If a star also has a planet, then the orbiting planet will tug the star around a common center of mass. By measuring this varying reflex velocity in the stellar spectrum over time, the orbital parameters of a planetary companion can be derived. 

The magnitude of the radial velocity signal depends on several factors, including the mass of the star, the mass of the planet, the orbital period, the shape (eccentricity), and the orientation of the orbit. Since orbits that are oriented ``face-on'' are tangential to our line of site, they do not have a radial component and therefore cannot be detected with the RV method. Fortunately, face-on orbits are a statistically rare configuration. 

In the solar system, Jupiter induces a radial velocity of about 12 \ms\ in the Sun while the lower mass Earth only induces a velocity of 0.09 \ms. If observed with very high spectral resolution, one pixel on the detector spans about 500 \ms, so these radial velocities would only shift the solar spectrum by $0.024$ or $0.00018$ pixels, for Jupiter and the Earth, respectively. Further complicating the detection, these tiny shifts are merely the semi-amplitudes of nearly sinusoidal variations with periods of about 12 years for Jupiter and 1 year for the Earth. Detecting such a tiny sub-pixel shift in stellar absorption features is non-trivial. The state-of-the-art Doppler precision for the past decade has been about 1 \ms\ \citep{fischer16}. This is sufficient to detect Jupiter (with 12 years of observations), but precludes the detection of Earth analogs around Sun-like stars. Because the RV amplitude increases with decreasing stellar mass, some Earth mass planets have been detected around stars that are lower in mass than the Sun.
Figure \ref{fig:exoplans} shows the velocity amplitudes and orbital periods of exoplanets detected over the past 25 years.
 }

The RV error budget includes instrumental errors, photon statistics (shot noise), and velocities from within the photosphere of the star that introduce scatter to the Keplerian velocities \citep{halverson16, dumusque17, blackman2020}. The EXtreme PREcision Spectrometer (EXPRES) \citep{jurgenson16, blackman2020, petersburg2020} is a newly commissioned instrument that was designed to significantly reduce instrumental errors. The primary goal of the EXPRES instrument is to provide higher fidelity data (high signal-to-noise with reduced instrumental errors) and the instrument has demonstrated intrinsic instrumental measurement precision better than 0.1 \ms\ \citep{blackman2020}. The next critical step for reaching Earth-detecting precision is the development of statistical techniques that estimate velocities with high precision and are less sensitive to photospheric velocities \citep{dumusque14, rajpaul15, dumusque17, davis17, rajpaul20}.

The traditional cross correlation function (CCF) \citep{baranne96} has long been used to measure Doppler shifts in stellar spectra by minimizing a weighted dot product between the observed spectrum and a template \citep{pepe02}. Various template matching algorithms have also been \changes{developed}, which minimize the (interpolated) sum of squared differences between the spectrum and a template spectrum using the Doppler shift as a free parameter \citep{anglada-escude12, astudillo15}. A variant of the template matching approach assumes the Doppler shift is small and estimates the derivative of the spectrum from the template \citep{bouchy01, dumusque18}. The EXPRES analysis pipeline has implemented the CCF method, as well as a higher precision Forward-Modeling (FM) code that makes use of a very high signal-to-noise (S/N) stellar template to model a Doppler shift in every 2-\AA\ segment of the observed spectrum \citep{petersburg2020}.

\changes{The new method we propose for estimating the RV is designed to work well in the small RV regime typical of orbiting exoplanets. 
Additionally, the proposed method is developed to generalize well to different types of stars because the modeling is carried out on the spectra observed for an individual star, and it does not require a pre-specified template. The only interpolation that takes place in the proposed method is on a high S/N, oversampled, template spectrum. Compared to the approach of \cite{anglada-escude12} which requires interpolation of every (low S/N) observed spectrum, the numerical error introduced through interpolation is likely reduced in the new proposed method.  Perhaps most importantly, the new method simplifies the RV estimation process to simple linear regression. This allows the method to easily account for the heteroskedastic noise in spectra. Furthermore, this simplification allows for straight-forward statistical inference on the estimated RV without making assumptions regarding the validity of propagation error or other approximate estimates of the standard error.}

\changes{The proposed Hermite-Gaussian Radial Velocity (HGRV) estimation method makes use of the well-known Hermite-Gaussian functions. These functions have been used extensively in modeling with Schrodinger's Equation \citep{marhic78, dai16}, as well as in fitting emission lines in galaxy spectroscopy \citep{riffel10}. The key contribution of this paper is that shifts of spectral lines between two spectra (e.g., due to a Doppler shift) can be well estimated with the first Hermite-Gaussian function fitted to the difference spectrum. 

The use of the Hermite-Gaussian functions is partially a consequence of the method's assumption that absorption features are Gaussian shaped. While the traditional CCF approach is designed to not depend on the individual shapes of absorption features by its use of a mask, and the template matching approaches take full account of absorption feature shapes, the HGRV approach can be thought of as between these two extremes in that it assumes the features are Gaussian-shaped.  It is important to note that large optical depth, rotational broadening, collisional broadening, stellar activity, and other astrophysical effects can cause absorption features to depart from a Gaussian shape. (The model misspecification due to this Gaussian-shape assumption is explored in Section \ref{model_misspecification}.) }

In Section 2 we introduce the data commonly used in the RV method, namely stellar spectra. We also propose an algorithm for finding absorption features in the spectrum that will be used in the HGRV method. \changes{Section 3 includes details of the proposed HGRV method, and simulation study results are discussed in Section 4. Section 5 then applies the method to} recently collected data of 51 Pegasi by EXPRES. A discussion is provided in Section 6 and we conclude in Section 7.

\section{Absorption Feature Finding Algorithm} \label{aff_section}

\begin{figure}[h!]
\centering
\includegraphics[scale=0.8]{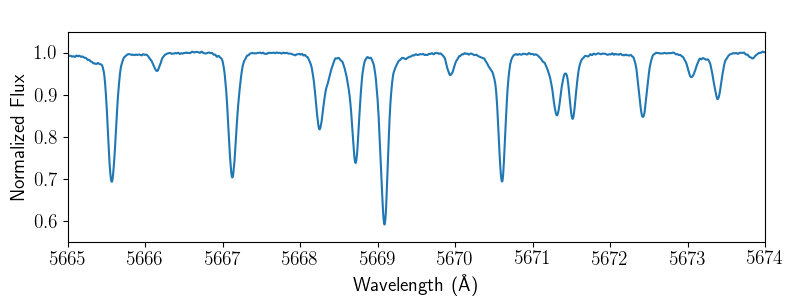}
\caption{A subset of the NSO spectrum of the Sun between $5665$ and $5674\ \AA$.}
\label{nsoftrs0}
\end{figure}

A small section of the Sun's spectrum, as collected by the National Solar Observatory (NSO) \citep{rimmele98}, is shown in Figure \ref{nsoftrs0}. \changes{In general, such a spectrum gives a representation of the relative brightness (hereafter referred to as normalized flux) as a function of wavelength. The narrow dips in the normalized flux are spectral absorption features which have variable intensity and frequent blending with neighboring features. In the (unrealistic) situation of these absorption features not being present, the remaining spectrum is referred to as the continuum.

The astrophysical blackbody effect \citep{planck01}, together with the instrumental effect often referred to as the blaze function, lead to a continuum that is not flat in the raw spectrum. However, various normalization techniques have been developed to correct for these effects \citep{xu19, petersburg2020}. A spectrum where the continuum has been normalized by dividing out the instrumental blaze and the blackbody curve is hereafter referred to as a normalized spectrum. Figure \ref{nsoftrs0} is an example of such a normalized spectrum.}

 \changes{We define the template spectrum of a star, $\tau$, to be its noiseless, normalized spectrum with no instrumental or astrophysical effects (e.g., activity such as spots). 
 Furthermore, we define the difference flux to be the difference between a single observed normalized spectrum and this template. An important characteristic of the HGRV method is that, rather than modeling a Doppler shift in the spectrum as a change in the explanatory variable (wavelength) as the CCF method does, we can model the difference in normalized flux caused by the Doppler shift. This characteristic is present in various other RV detection methods \citep{bouchy01, rajpaul20}, but it is implemented rather differently with our proposed method.}

Since a Doppler shift only rescales the wavelength axis, there is \changes{little} RV information in the normalized continuum. Most of the information for small Doppler shifts comes from the slopes of spectral lines, so identification of the absorption features in a given spectrum is the first step for the HGRV method. 

\changes{The locations, depths, and degree of blending of absorption features depend on the stellar parameters and chemical composition of the star and, therefore, vary from star to star. 
The HGRV method involves modeling individual absorption features so an algorithm is needed that not only identifies the central wavelength at which each feature occurs, but also the wavelength bounds that contain the feature. Were all absorption features to be well-separated, these wavelength bounds would nearly be symmetric about the central wavelengths with a nearly-constant width. However, since blends are very common, this is not the case in practice.}

\changes{Designing the HGRV method to generalize across stars motivates the use of an algorithm for identifying absorption feature wavelength bounds in a way that can adapt to different spectra. The proposed absorption feature finding algorithm is a statistically-motivated heuristic algorithm. The overarching goal is to find wavelength windows of absorption features, not to perform any statistical inference on them. } 

\changes{The algorithm has two main sequential steps: (i) identify local minima that are likely to be absorption lines and (ii) proceed outward from each local minimum until the normalized flux flattens out. This algorithm is presented in Algorithm \ref{aff_algorithm} and requires three tuning parameters: a wavelength window size $m$ in units of pixel count, and significance levels $\alpha,\ \eta$ where $\eta \geq \alpha$. For a more thorough motivation of this algorithm, as well as a more detailed overview of the steps involved, see Appendix \ref{appendix_aff}.}

\begin{algorithm}[h!]
\SetAlgoLined
\DontPrintSemicolon
\KwData{ordered wavelengths $\Lambda = \left( x_{0}, x_{1}, ... , x_{n} \right)$ and corresponding flux values $\tau = \left( \tau_{0}, \tau_{1}, ... , \tau_{n} \right)$}
Initialize tuning parameters $m \in \mathbb{N}$, $\alpha \in (0,1)$, and $\eta \in (\alpha, 1)$ \;
\For{$x_{i} \in \Lambda$}{
set $\Lambda_{l,i} = \left( x_{i-m+1}, x_{i-m+2}, ..., x_{i} \right)^{T}$, $\Lambda_{r,i} = \left( x_{i}, x_{i+1}, ..., x_{i+m-1} \right)^{T}$, $\tau_{l,i} = \left( \tau_{i-m+1}, \tau_{i-m+2}, ..., \tau_{i} \right)^{T}$, and $\tau_{r,i} = \left( \tau_{i}, \tau_{i+1}, ..., \tau_{i+m-1} \right)^{T}$\;
model $\tau_{l,i} = \beta_{0,l} \mathds{1}_{m} + \beta_{1,l} \Lambda_{l,i} + \varepsilon$ and $\tau_{r,i} = \beta_{0,r} \mathds{1}_{m} + \beta_{1,r} \Lambda_{r,i} + \varepsilon '$ where $\varepsilon, \varepsilon ' \sim N\left(0, \varsigma^{2} I_{m}\right)$ and $\mathds{1}_{m} = (1,1,...,1)^{T}$ with length $m$ \;
get p-values $p_{l,i}$ for testing $\beta_{1,l} = 0$ against $\beta_{1,l} < 0$ and $p_{r,i}$ for testing $\beta_{1,r} = 0$ against $\beta_{1,r} > 0$ \;
}
Initialize index $j = m$ and upperbound $u = 0$ \;
\While{$j \leq \  \mathrm{length}(\Lambda) - m + 1$}{
\eIf{$p_{l, j} < \alpha / 2 \ \mathrm{and} \ p_{r, j} < \alpha / 2$}{
set $k_{\mathrm{max}} = \mathrm{max} \left\lbrace k \in \left\lbrace u, u+1,...,j \right\rbrace : p_{l,k} \geq \eta \right\rbrace$ \;
set $k_{\mathrm{min}} = \mathrm{min} \left\lbrace k \in \left\lbrace j, j+1, ..., \mathrm{length}(\Lambda) \right\rbrace : p_{r,k} \geq \eta \right\rbrace$ \;
save $\left( \dfrac{x_{k_{\mathrm{max}}} + x_{k_{\mathrm{max}} - m}}{2}, \dfrac{x_{k_{\mathrm{min}}} + x_{k_{\mathrm{min}}  + m}}{2} \right)$ as absorption feature wavelength bounds \;
$j \leftarrow \lfloor \left( k_{\mathrm{min}} + m/2 \right) \rfloor$ \; 
$u \leftarrow j$ \;
}{
$j \leftarrow j+1$ \;
}
}
\caption{Absorption Feature Finder}
\label{aff_algorithm}
\end{algorithm}

\begin{figure}[h!]
\centering
\includegraphics[scale=0.8]{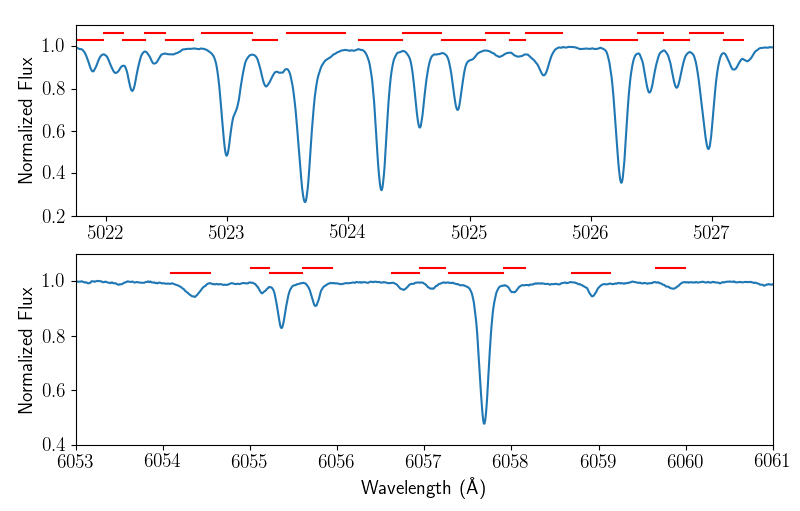}
\caption{Results of using Algorithm \ref{aff_algorithm} on the NSO Spectrum. Red horizontal lines show the wavelength windows found to correspond to individual absorption features.}
\label{nsoftrs1}
\end{figure}

Algorithm \ref{aff_algorithm} was empirically evaluated using the NSO spectrum. After the step-by-step optimization of the three tuning parameters described in Appendix \ref{appendix_aff}, we found that $m = 25$, $\alpha = 0.01$, and $\eta = 0.05$ found the most absorption features\changes{. Furthermore, we visually-identified no} false positives remaining after eliminating features with a line depth less than $0.015$. A subset of the absorption features found in the NSO spectrum are shown in Figure \ref{nsoftrs1}.

\changes{To estimate the false-positive rate of this algorithm, we considered the NSO spectrum between $5000$ and $6000$ \AA\ and replaced the normalized flux axis with a flat $500$ S/N simulation 20 independent times.  See Sections \ref{template_est} and \ref{template_est_sim} for details on how we estimate a template spectrum with this level of S/N, and which we use in Algorithm \ref{aff_algorithm}. Applying Algorithm \ref{aff_algorithm} to these simulations with parameters $m = 25$, $\alpha = 0.01$, and $\eta = 0.05$ gave a total of $55$ detected features. Since the spectra did not have any absorption features, this approximates the false positive rate as $1$ absorption feature per $363$ \AA. Additionally, the line depths of these $55$ false features had mean $0.0046$, standard deviation $0.0018$, and maximum $0.0098$ so that all the false lines would be eliminated with the minimum line depth parameter set to $0.015$.} \changes{Note that for spectra with either different S/N or resolution $m$, $\alpha$, $\eta$, and the minimum line depth may need to be adjusted (e.g., a lower S/N or resolution may need higher significance levels or a higher minimum line depth). We recommend setting $m$ to be approximately $25 \times \dfrac{R}{2 \times 10^{6}}$ where $R$ is the resolution of the spectrum, and the minimum line depth to be approximately $0.015 \times \dfrac{500}{\mathrm{S/N}}$. For details on this recommendation see Appendix \ref{appendix_aff}.}

\changes{In addition we applied Algorithm \ref{aff_algorithm} directly to the NSO spectrum between $5000$ and $6000$ \AA. We found that the wavelength bounds given by the algorithm contained $64.3\%$ of the spectrum, but accounted for $97.7\%$ of the mean squared deviations from $1.0$ of the normalized flux. The remaining $2.3\%$ was mostly due to occasional absorption features whose overall shape due to line blends seemed to contribute to the algorithm missing them. 
For some additional plots associated with these results, see Appendix \ref{appendix_aff}.}

The proposed algorithm \changes{may have difficulty distinguishing two
spectral lines that are strongly blended together because the slope of the normalized flux may not flatten out between the two lines.} 
Depending on the S/N of the spectrum, \changes{it may not be able to find small features} as the noise would reduce the statistical significance of the left and right slopes. 
The lower the S/N is, the narrower the wavelength bounds will be for each detected absorption feature. This is because as we move outwards from the central wavelength of a feature, the slope eventually decreases in magnitude and becomes statistically insignificant sooner in the presence of more noise. \changes{We find that as long as the spectrum has a S/N above $500$ the results of our algorithm are stable whether or not one accounts for the heteroskedastic nature of the noise. We use the estimated template spectrum (described in Section \ref{template_est}) in Algorithm \ref{aff_algorithm}, and demonstrate in Section \ref{template_est_sim} that the template has a S/N above $500$ as long as there are at least $11$ observed spectra provided.}

\section{Hermite-Gaussian RV Method}

We now introduce the HGRV method by first considering the \changes{difference between a Gaussian and a multiplicative shift of it}. We introduce a theorem that quantifies the approximation error made by using only the first-degree Hermite-Gaussian function to model this difference, and provide the proof through four lemmas (the proofs of which \changes{can be found} in Appendix \ref{appendix_proofs}). We then show that, in the context of stellar spectroscopy, this approximation error is small and the coefficient of the first-degree Hermite-Gaussian function is nearly a constant multiple of the RV. This allows us to extend to the case of multiple absorption features and reduce the problem of estimating the Doppler shift in a spectrum to linear regression.

\subsection{Mathematics of a Doppler-shifted Gaussian}

If $x$ represents the wavelength of light and $f(x)$ represents the normalized flux of light at that wavelength, then the normalized flux of Doppler-shifted light is represented mathematically as $f(\xi x)$ where \changes{$\dfrac{1}{\xi}$} is referred to as the Doppler factor \citep{doppler42}. \changes{In special relativity, $\xi$ is given by 
\begin{equation}
\xi = \dfrac{1 + v_{r}/c}{\sqrt{1 - (v/c)^{2}}} \label{dopplershift}
\end{equation}
where $c$ is the speed of light \citep{einstein05}, $v$ is the absolute speed of the source, and $v_{r}$ is the velocity along the line of site of the observer. While the Earth's rotation and revolution around the solar system barycenter often lead to relativistic effects, these motions are well understood and can be corrected for with high precision \citep{wright14, blackman17, blackman2020}. Furthermore, the velocity due only to orbiting exoplanets is well below the speed of light. Therefore, under the assumption that the barycentric corrections are applied accurately and $v \ll c$, $\xi$ can be well approximated with the classical formula
\begin{equation}
    \xi = 1 + \dfrac{v_{r}}{c} \label{dopplershift_classical} .
\end{equation}

}

Consider the effect of a Doppler shift when $f(x)$ is a Gaussian like many of the inverted absorption features in a spectrum \citep{gray05}. To model this we propose the Hermite-Gaussian functions, $\psi_{n} (x)$, defined as
\begin{equation}
\psi_n(x) = \dfrac{1}{\sqrt{2^n n!\sqrt{\pi}}} H_n(x) e^{-(x^2)/2} \label{hermgauss_def}
\end{equation}
where $H_n(x)$ represents the $n$'th degree (physicist's) Hermite polynomial which can be written in closed form as
\begin{eqnarray}
H_{k}(s) = k! \sum\limits_{m=0}^{\lfloor k/2 \rfloor} \dfrac{(-1)^{m}}{m! (k-2m)!} (2s)^{k-2m}
\label{hermitepoly}
\end{eqnarray}
with $\lfloor a \rfloor$ representing the floor function that returns the largest integer less than or equal to the real number $a$ \citep{lanczos38}.

\begin{figure}[h!]
\centering
\includegraphics[scale=0.5]{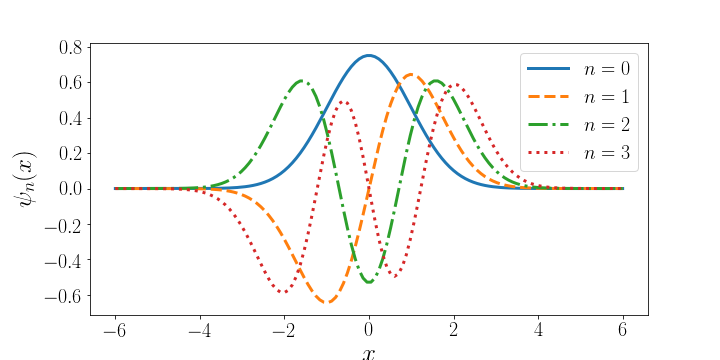}
\caption{The first $4$ Hermite-Gaussian functions given by Equation \eqref{hermgauss_def}.}
\label{hermgauss_functions}
\end{figure}

An illustration of the first four Hermite-Gaussian functions is shown in Figure \ref{hermgauss_functions}.

According to \cite{johnston14}, 
\begin{equation}
\int_{-\infty}^{\infty} H_n(x) H_m(x) e^{-x^2} dx = \sqrt{\pi} 2^n n! \mathds{1}\{m = n\} \label{hermpol_innerproduct}
\end{equation}
is a well known fact about the Hermite polynomials, where $\mathds{1}\{ A \}$ represents the indicator function of the event $A$ (which is equivalent to the Kronecker delta function).

Therefore, we have by combining equations \eqref{hermgauss_def} and \eqref{hermpol_innerproduct} that 
\begin{equation}
\int_{-\infty}^{\infty} \psi_n(x) \psi_m(x) dx = \mathds{1}\{m = n\} \label{hermgauss_innerproduct}  .
\end{equation}
Furthermore, one can show that the set of Hermite-Gaussian functions forms a complete orthonormal basis of the set of all square-integrable real-valued functions, $L^2(\mathbb{R})$ \citep{johnston14}. One can also generalize the definition of the Hermite-Gaussian functions to have a general \changes{location}, $\mu$, and \changes{scale}, $\sigma$:
\begin{equation}
\psi_n(x; \mu, \sigma) = \dfrac{1}{\sqrt{\sigma 2^n n!\sqrt{\pi}}} H_n\left(\dfrac{x - \mu}{\sigma}\right) e^{-\dfrac{(x - \mu)^2}{2 \sigma^2}} \label{genhermgauss_def}  .
\end{equation}
By a simple change of variables, one can show that the set of generalized Hermite-Gaussian functions, $\psi_n(x; \mu, \sigma)$, also forms a complete orthonormal basis of $L^2(\mathbb{R})$ for any $\mu \in \mathbb{R}$ and any $\sigma \in \mathbb{R}^{+}$, the positive real numbers. Therefore, for such an $L^2(\mathbb{R})$ function $g$, we can decompose it as
\begin{equation}
g(x) = \sum\limits_{n=0}^{\infty} c_{n} \psi_n(x ; \mu, \sigma) \label{hermgauss_decomposition}  .
\end{equation}
In this instance let $f(x)$ be a Gaussian with center $\mu$ and width $\sigma$, and let $g(x; \xi) = f(x) - f(\xi x)$ be the difference between $f(x)$ and its Doppler-shifted version. Decomposing this $g(x; \xi)$ as in Equation \eqref{hermgauss_decomposition}, we have Theorem \ref{maintheorem}, giving the approximation error when only $n=1$ is used.

\newpage
\begin{theorem}
For any $\sigma \in \mathbb{R}^{+}$ and any $\mu, \xi \in \mathbb{R}$ and $g(x; \xi) = e^{-\dfrac{(x-\mu)^{2}}{2 \sigma^{2}}} - e^{-\dfrac{(\xi x-\mu)^{2}}{2 \sigma^{2}}}$ decomposed in the Hermite-Gaussian basis as $g(x; \xi) = \sum\limits_{n=0}^{\infty} c_{n}(\xi) \psi_{n}(x ; \mu, \sigma)$, \\
%\begin{center} 
\begin{equation}
\lim\limits_{\xi \rightarrow 1} \dfrac{\int_{-\infty}^{\infty} \left( g(x; \xi) - c_{1}(\xi) \psi_{1}(x ; \mu, \sigma)\right)^{2} dx}{\int_{-\infty}^{\infty} \left( g(x; \xi) \right)^{2} dx} = \dfrac{1}{1 + \dfrac{2 \mu^2}{3 \sigma^2}}. 
\end{equation}
%\end{center}
\label{maintheorem}
\end{theorem}
Before proving Theorem \ref{maintheorem}, we interpret it in the context of stellar spectroscopy. It is well known that many absorption features in the spectrum of a star are described by the Voigt profile \citep{ciurylo98, gray05}, which is well approximated by a Gaussian for many absorption features in stellar spectra. It is also the case that the central wavelength, $\mu_x$, is significantly larger than the width, $\sigma_x$, for each of these features. As an example, a typical wavelength in the visible spectrum is $5000$ \AA, and the largest features near this wavelength have a width that is upper-bounded by $0.5$ \AA; the maximum width of absorption features detected between $4700$ \AA\ and $5300$ \AA\ by Algorithm \ref{aff_algorithm} for the data collected from 51 Pegasi by EXPRES was $0.366$ \AA\ with the $88$'th quantile being $0.1$ \AA\ (more details to come in Section 5). For a feature with center $5000$ \AA \ and width $0.5$ \AA, the limit in Theorem \ref{maintheorem} becomes $1.5\times 10^{-8}$. Therefore the theorem implies that as $\xi$ approaches $1$ (i.e. at small values of RV), the proportion of the difference, $g(x; \xi)$, that remains to be modeled after using only $\psi_{1}$ \textit{with the same width and center as the original Gaussian} is nearly zero. In other words, Doppler shifting a Gaussian absorption feature at a small RV is approximately the same as adding a constant multiple of $\psi_{1}$ \changes{(which is a scalar multiple of the Gaussian's derivative)} to the feature.

Some of the RV detection algorithms, such as the template matching method described in \cite{bouchy01}, attempt to model a Doppler shift by approximating the derivative of absorption features with a high S/N template spectrum. \changes{They then use a wavelength multiple of this derivative to create a nonlinear model of a Doppler shift with parameters to be fitted.} At high wavelength values, though, multiplication of a narrow wavelength window is nearly the same as an additive shift. In fact, if the Doppler shift were additive, the limit in Theorem \ref{maintheorem} would be $0$. \changes{Furthermore, an additive shift removes the nonlinearity in the Doppler shift model.} \changes{While this idea is not new \citep{butler96},} the approximation error of this has remained unknown. Therefore, Theorem \ref{maintheorem} takes account of the multiplicative nature of the Doppler shift, giving the value of this approximation error for assuming the shift to be additive at the limit of low values of RV.

To answer the question of how small an RV is small enough for this to be valid, we first state some Lemmas that solve for the coefficients in the decomposition shown in Equation \eqref{hermgauss_decomposition} with $g(x; \xi)$ as defined in Theorem \ref{maintheorem}. Lemma \ref{intk_lemma} gives a useful recursive relationship of an integral quantity that arises in solving the coefficients.

\vspace{1cm}
\begin{lemma}
For $I_{k} (a, b, c) := \int_{-\infty}^{\infty} u^{k} e^{-\left( a u^{2} + b u + c \right)} du$ where $a > 0$, we have that \\ \begin{equation}
I_{0} (a, b, c) = \sqrt{\dfrac{\pi}{a}} e^{\left( \dfrac{b^{2}}{4a} - c \right)},
\end{equation} \begin{equation}
I_{1} (a, b, c) = - \dfrac{\sqrt{\pi} b}{2 a^{3/2}} e^{\left( \dfrac{b^{2}}{4a} - c \right)},
\end{equation}  \begin{equation}
\mathrm{and\ for\ all\ } k \geq 2,\ I_{k}(a,b,c) = -\dfrac{b}{2a} I_{k-1}(a,b,c) + \dfrac{k-1}{2a} I_{k-2}(a,b,c).
\end{equation} \label{intk_lemma}
\end{lemma}

\begin{figure}[h!]
\centering
\includegraphics[scale=0.68]{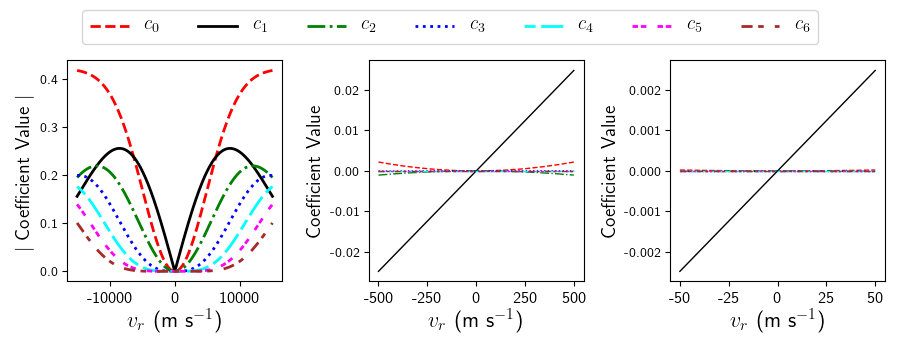}
\caption{The coefficient solutions that result from modeling a Doppler-shifted Gaussian with the Hermite-Gaussian basis are plotted here as a function of $v_{r}$. The left panel has the absolute value of the coefficients on the vertical axis and illustrates that at low values of $v_{r}$, $c_{1}$ is the dominating coefficient. The middle and right panels show the exact coefficient value and illustrate that at low values of $v_{r}$, $c_{1}$ is nearly a constant multiple of it. Only the zero'th up to the sixth coefficients are shown. The Gaussian here has the parameters of $\mu = 5000$ and $\sigma = 0.1$ which is meant to represent a typical absorption feature in a stellar spectrum.}
\label{coef_vs_rv}
\end{figure}

Using $I_{k}(a,b,c)$ as defined in Lemma \ref{intk_lemma}, Lemma \ref{ck_lemma} gives the mathematical solution for the coefficients.
\begin{lemma}
For $g(x; \xi) = e^{-\dfrac{(x-\mu)^{2}}{2 \sigma^{2}}} - e^{-\dfrac{(\xi x-\mu)^{2}}{2 \sigma^{2}}}$ decomposed as $g(x; \xi) = \sum\limits_{n=0}^{\infty} c_{n}(\xi) \psi_{n}(x ; \mu, \sigma)$, and $I_{k} (a, b, c)$ as defined in Lemma \ref{intk_lemma}, we have that for $\varepsilon = \xi -1$ \\ \begin{equation}
c_{0}(\varepsilon) = \sqrt{\sigma \sqrt{\pi}} - \dfrac{1}{\sqrt{\sigma \sqrt{\pi}}} I_{0}\left( \dfrac{1 + \varepsilon + \dfrac{\varepsilon^{2}}{2}}{\sigma^{2}}, -\dfrac{2\mu + \varepsilon \mu}{\sigma^{2}}, \left( \dfrac{\mu}{\sigma} \right)^{2} \right),
\end{equation}
and for all $k \geq 1$, 
\begin{equation}
c_{k}(\varepsilon) = - \sqrt{\dfrac{\sigma k! 2^{k}}{\sqrt{\pi}}} \sum\limits_{m=0}^{\left \lfloor \dfrac{k}{2} \right \rfloor} \dfrac{(-1)^{m}}{4^{m} m! (k-2m)!}I_{k-2m} \left( 1 + \varepsilon + \dfrac{\varepsilon^{2}}{2} , \dfrac{\varepsilon \mu}{\sigma}(1+\varepsilon), \dfrac{1}{2}\left( \dfrac{\varepsilon \mu}{\sigma} \right)^{2} \right).
\end{equation} \label{ck_lemma}
\end{lemma}
Using Lemmas \ref{intk_lemma} and \ref{ck_lemma} we numerically calculate the first seven coefficients as a function of RV and illustrate the results in Figure \ref{coef_vs_rv}. It is not hard to notice that all the coefficients go to $0$ as the RV goes to $0$. This is because with no RV, $g(x; \xi)$ as defined in Theorem \ref{maintheorem} is the zero-function. More importantly, though, Figure \ref{coef_vs_rv} illustrates that as the RV approaches zero, the dominating coefficient is $c_{1}$.

When $v_{r}$ has a magnitude below 100 \ms\ it appears that all other coefficients besides $c_{1}$ are negligible, with $c_{0}$ and $c_{2}$ being the only possible exceptions. Furthermore, at velocities with a magnitude below 500 \ms, $c_{1}$ is approximately linear as a function of $v_{r}$. Since Figure \ref{fig:exoplans} illustrates that a considerable number of currently known exoplanets exert a RV on their host star with a semi-amplitude less than 100 \ms, which is especially true for Earth-like exoplanets, it suggests that it is not unreasonable to ignore all Hermite-Gaussian coefficients besides $c_{1}$ in modeling a Gaussian absorption feature that is Doppler-shifted due to an exoplanet.

Now that we have the coefficient solutions, and have a sense that $c_{1}$ is the most dominant coefficient at values of RV that are of interest, we calculate the approximation error made by ignoring all other coefficients. To do so, we introduce a new quantity that we refer to as the standardized approximation error, which appears in Theorem \ref{maintheorem}. For a function $\varphi$ approximated by the function $\phi$, define the standardized approximation error $D(\phi || \varphi)$ as
\begin{eqnarray}
D(\phi || \varphi) = \dfrac{\int_{-\infty}^{\infty} \left( \varphi(x) - \phi(x) \right)^{2} dx}{\int_{-\infty}^{\infty} \varphi(x)^{2} dx} \label{std_approx_err} \ .
\end{eqnarray}

In a sense, $D(\phi || \varphi)$ gives the proportion of the \changes{squared} function $\varphi$ that remains to be modeled after approximating with $\phi$. In our case we consider $D\left( g(x; \xi) || c_{1}(\xi) \psi_{1}(x ; \mu, \sigma) \right)$.\footnote{Since $g(x; \xi)$ approaches the zero function as $\xi \rightarrow 1$, and for any $k \geq 0$ $c_{k}(\xi) \rightarrow 0$ as $\xi \rightarrow 1$, the ordinary approximation error of using any individual $k$ would approach $0$. This would tell us nothing about the relative magnitudes of the Hermite-Gaussian coefficients. The denominator of $D\left( g(x; \xi) || c_{1}(\xi) \psi_{1}(x ; \mu, \sigma) \right)$ adjusts for this by standardizing the quantity.} Lemmas \ref{ratio_lemma} and \ref{final_lemma} help us solve for the limit as $\xi$ approaches $1$ (i.e. as $v_{r}$ approaches $0$).
\begin{lemma}
For $g(x; \xi) = e^{-\dfrac{(x-\mu)^{2}}{2 \sigma^{2}}} - e^{-\dfrac{(\xi x-\mu)^{2}}{2 \sigma^{2}}}$ decomposed as \\ $g(x; \xi) = \sum\limits_{n=0}^{\infty} c_{n}(\xi) \psi_{n}(x ; \mu, \sigma)$, we have that 
\begin{equation}
D\left( g(x; \xi) || c_{1}(\xi) \psi_{1}(x ; \mu, \sigma) \right) = 1 - \dfrac{c_{1}^{2}(\xi)}{\int_{-\infty}^{\infty} \left( g(x; \xi) \right)^{2} dx}.
\end{equation}
\label{ratio_lemma}
\end{lemma}
\begin{lemma}
$\lim\limits_{\xi \rightarrow 1} \dfrac{c_{1}^{2}(\xi)}{\int_{-\infty}^{\infty} \left( g(x; \xi) \right)^{2} dx} = \dfrac{1}{1 + \dfrac{3\sigma^{2}}{2\mu^{2}}}$. \label{final_lemma}
\end{lemma}
Combining Lemmas \ref{ratio_lemma} and \ref{final_lemma} completes the proof of Theorem \ref{maintheorem}. (See Appendix \ref{appendix_proofs} for a more detailed proof of each.)\footnote{It is worth noting that in the proof of Lemma \ref{final_lemma}, L'hopital's rule must be applied twice. And since $c_{1}^{2}(\xi)$ is essentially an integral, one would naturally suggest that the proof could be simplified by interchanging two derivatives and the limit with the integration in both the numerator and denominator. However, it can be shown that this results in the limit of Lemma \ref{final_lemma} incorrectly being $1$. Therefore, this is an instance in which this interchange is not mathematically valid and cannot be used to simplify the proof.}

\begin{figure}[h!]
\centering
\includegraphics[scale=0.68]{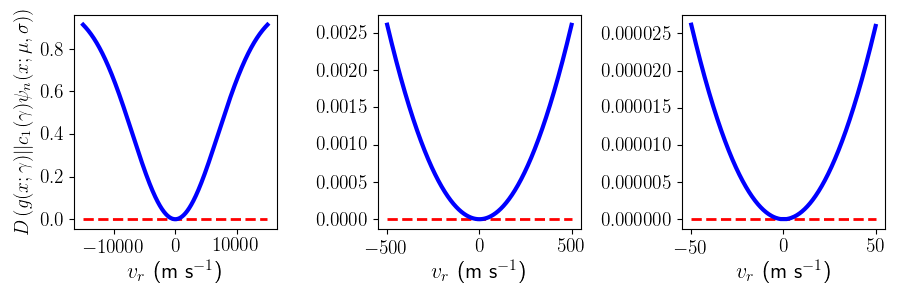}
\caption{The standardized approximation error $D\left( g(x; \xi) || c_{1}(\xi) \psi_{n}(x ; \mu, \sigma) \right)$ in Theorem \ref{maintheorem} as a function of $v_{r}$ with parameters $\mu = 5000$ and $\sigma = 0.1$ is plotted in bold. The limit is also shown in the horizontal red dashed line.}
\label{approx_err}
\end{figure}

Theorem \ref{maintheorem} does not explicitly give a rate at which the standardized approximation error approaches its limit. But by using Lemma \ref{ratio_lemma} and Equation \eqref{denom_eqn} from the proof of Lemma \ref{final_lemma} in Appendix \ref{appendix_proofs}, we illustrate the rate with Figure \ref{approx_err}. Note that the standardized approximation error shown here is bounded between $0$ and $1$, and that the limit is actually non-zero. Figure \ref{approx_err} illustrates that as $\xi \rightarrow 1$, $D\left( g(x; \xi) || c_{1}(\xi) \psi_{n}(x ; \mu, \sigma) \right)$ approaches its limit quadratically and that \changes{when $v_{r} < 50$} \ms, the standardized approximation error is less than $2.5\times 10^{-5}$ away from the limiting value.

\subsection{RV Estimation Method}
Theorem \ref{maintheorem} suggests a natural new method for detecting a Doppler shift in the spectrum of a star.  As long as the magnitude of $v_{r}$ is small enough, the absorption feature is approximately Gaussian, and the ratio $\mu/\sigma$ for the feature is large enough, we can do a least-squares fitting of the first-degree Hermite-Gaussian function to the difference between a template spectrum and a Doppler-shifted spectrum and map the fitted coefficient to a RV. As illustrated in Figure \ref{coef_vs_rv}, $c_{1}$ at low values of $v_{r}$ is directly proportional to $v_{r}$. 

According to Lemma \ref{ck_lemma}, $c_{1}(\varepsilon) = \dfrac{\sqrt{\sqrt{\pi}}}{\sqrt{2\sigma}} \varepsilon \mu (1 + \varepsilon) \tilde{h}(\varepsilon)$, and $\lim\limits_{\varepsilon \rightarrow 0} \dfrac{\partial}{\partial \varepsilon} c_{1} (\varepsilon) = \dfrac{\mu\sqrt{\sqrt{\pi}}}{\sqrt{2\sigma}}$.

Furthermore, using Equation \eqref{dopplershift_classical} with $\varepsilon = \xi -1$, we have that the mapping from $\varepsilon$ to RV is \changes{$v_{r}(\varepsilon) = c\varepsilon$ and $\lim\limits_{\varepsilon \rightarrow 0} \dfrac{\partial}{\partial \varepsilon} v_{r}(\varepsilon) = c$}. Hence, $\lim\limits_{\varepsilon \rightarrow 0} \dfrac{\partial}{\partial v_{r}} c_{1}\left(v_{r}(\varepsilon)\right) =  \dfrac{\mu\sqrt{\sqrt{\pi}}}{c\sqrt{2\sigma}} $
%\begin{eqnarray*}
%\lim\limits_{v \rightarrow 0} \dfrac{\partial}{\partial v} c_{1}(v) = \lim\limits_{\varepsilon \rightarrow 0} \dfrac{\partial}{\partial v} c_{1}\left(v(\varepsilon)\right) = \lim\limits_{\varepsilon \rightarrow 0} \dfrac{\dfrac{\partial}{\partial \varepsilon} c_{1}\left(v(\varepsilon)\right)}{\dfrac{\partial}{\partial \varepsilon} v(\varepsilon)} = - \dfrac{\mu\sqrt{\sqrt{\pi}}}{c\sqrt{2\sigma}} 
%\end{eqnarray*}
which is the desired proportionality constant. So the proportionality that is valid at low values of RV, $v_{r}$, is
\begin{equation}
c_{1} = \dfrac{\mu\sqrt{\sqrt{\pi}}}{c\sqrt{2\sigma}} v_{r} \label{c1_to_v} \ .
\end{equation}
The strongest assumption made when applying the theorem is that the absorption features are Gaussian shaped. \changes{Because} this may never be exactly true, \changes{we analyze this model misspecification further in Section \ref{model_misspecification} below.}

\subsection{Extension to multiple features}
Since a single absorption feature is unable to give a RV estimate that is precise enough, we need to use as many features in the spectrum as possible. Instead of fitting only a single first-degree Hermite-Gaussian function to the difference spectrum, we fit a sum of these functions to it. To construct this sum, we note that it must take into account the fact that differing absorption features will have different centers, widths, and depths. The generalized Hermite-Gaussian functions in Equation \eqref{genhermgauss_def} can take account of the different centers and widths. Furthermore, according to Equation \eqref{coef_equation} in the proof of Lemma \ref{ck_lemma}, Doppler-shifting a Gaussian with any amplitude simply multiplies the resulting coefficients by the same amplitude. In the case of stellar spectra, this amplitude is simply the line depth. Therefore, using Equation \eqref{c1_to_v}, the resulting model of the difference flux at pixel $i$, $y_{i}$, as a function of wavelength, $x_{i}$, to be fitted becomes 
\begin{eqnarray}
y_{i} = v_{r} \sum\limits_{j = 1}^{n} \dfrac{\sqrt{\sqrt{\pi}} d_{j} \mu_{j}}{c \sqrt{2 \sigma_{j}}} \psi_{1}\left( x_{i} ; \mu_{j}, \sigma_{j} \right) + \varepsilon_{i} \label{bigXvariable},
\end{eqnarray}
%\begin{eqnarray}
%y_{i} = \sum\limits_{j = 1}^{n} - \dfrac{d_{j} \mu_{j} \sqrt{\sqrt{\pi}}}{c \sqrt{2 \sigma_{j}}} v \psi_{1}\left( x_{i} ; \mu_{j}, \sigma_{j} \right) + \varepsilon_{i} = v \sum\limits_{j = 1}^{n} - \dfrac{\sqrt{\sqrt{\pi}} d_{j} \mu_{j}}{c \sqrt{2 \sigma_{j}}} \psi_{1}\left( x_{i} ; \mu_{j}, \sigma_{j} \right) + \varepsilon_{i} \label{bigXvariable},
%\end{eqnarray}
where the sum is over all $n$ absorption features, $d_{j}$ represents the line depth of the $j$'th feature, and each $\varepsilon_{i}$ is independent with expectation $0$.

\changes{In practice, we assume that $\varepsilon_{i}\ \sim\ \ N\left(0, \varrho_{i}^{2}\right)$ and is independent for each $i$. Many modern stellar spectra come with uncertainties for each pixel's normalized flux.\footnote{If these uncertainties are not provided, weights can be defined using the standard assumption that the raw flux is Poisson.  That is, the weights can be set to $w_{i} = \dfrac{\mathrm{cont}_{i}}{\hat{\tau}_{i}}$ where $\mathrm{cont}_{i}$ is the value of the raw continuum used for normalization at pixel $i$ and $\hat{\tau}_{i}$ is the value of the estimated template. 
}  This is particularly true for the normalized spectra from EXPRES that we analyze here. EXPRES estimates the uncertainty in each pixel by assuming the unnormalized flux is Poisson, estimating the red noise, and accounting for intrinsic effects of flat-fielding \citep{petersburg2020}. Therefore,  we assume that the provided uncertainties, $\hat{\varrho}_{i}$, are accurate estimates of each $\varrho_{i}$, and estimate $v_{r}$ in Equation \eqref{bigXvariable} through weighted least squares with weights $w_{i} = 1/\hat{\varrho}_{i}^{2}$.

}

To calculate the difference flux, $y_{i}$, at pixel $i$ we need a template spectrum. Here we use the \changes{estimated template} calculated from the set of observed spectra (see Section \ref{template_est} for more details). 

Since Equation \eqref{c1_to_v} approximately holds for $v_{r}\ <\ 500$ \ms, which well encompasses most exoplanets of interest, we have a new Hermite-Gaussian based Radial Velocity (HGRV) estimation method. For a spectrum of Gaussian absorption features, we can create a linear model of the difference spectrum due to a Doppler-shift as a function of the sum of $\psi_{1}$ functions as given by Equation \eqref{bigXvariable}, the coefficient of which is the RV. Therefore, we have reduced the Doppler shift estimation problem to linear regression with no intercept. This method does not include interpolation\footnote{Interpolation is, however, used later on a high S/N, oversampled estimate of the template spectrum to give it the same wavelength solution as each observed spectrum so that the difference flux can be calculated.}, treats neighboring pixels similarly, \changes{accounts for the hetroskedastic noise,} and easily allows for statistical inference.

\changes{
\subsection{Model Misspecification}\label{model_misspecification}

The HGRV method assumes that the shape of absorption features is Gaussian, which often does not hold exactly. Various reasons are understood to contribute to this: a line following the Voigt profile may have a non-negligible Lorentzian component, the line may be deep enough to depart from the Voigt profile, or there may be additional effects in the star's atmosphere that are not well-encompassed by current physical models. 

Since the HGRV method assumes Gaussian shaped absorption features, we now investigate the effects of applying it to non-Gaussian shaped features. We consider the absorption feature in the NSO spectrum between $5243.7$ and $5244.2$ \AA. This feature is shown in the left panel of Figure \ref{modelmisspec1}, along with its best-fit Gaussian. For $50$ equally spaced values of RV from $1$ to 100 \ms\ we Doppler shift this feature according to Equation \eqref{dopplershift_classical}, use cubic splines to interpolate back to the original wavelength solution \citep{meszaros13}, and fit the difference flux with the HGRV model from Equation \eqref{bigXvariable} (with $n=1$ and $d$, $\mu$, and $\sigma$ as the estimated parameters from the best-fit Gaussian). The ratio between the estimated and true RV is shown in the right panel of Figure \ref{modelmisspec1}.

\begin{figure}[h!]
\centering
\includegraphics[scale=0.68]{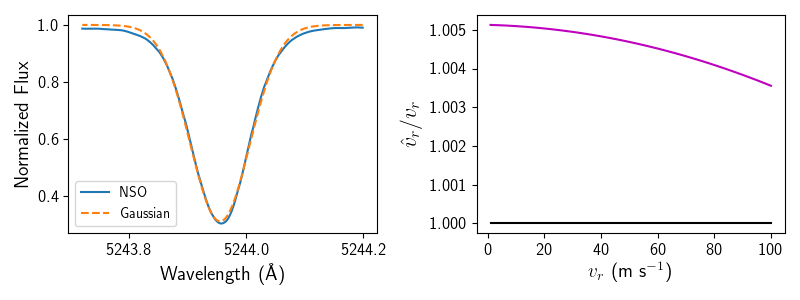}
\caption{Results for analyzing the effects of misspecifying the model of the absorption feature in the NSO spectrum between $5243.7$ and $5244.2$ \AA\ as a Gaussian. The left panel shows the feature in solid blue and the best-fit Gaussian in dashed orange. The right panel shows the ratio of the RV estimated with Equation \eqref{bigXvariable} $\hat{v}_{r}$ (with $n=1$) and the true RV, $v_{r}$.}
\label{modelmisspec1}
\end{figure}

Figure \ref{modelmisspec1} illustrates that for this particular absorption feature, the HGRV method slightly overestimates the RV. For example, if the true RV is $1$ \ms, this bias would be approximately $0.5$ cm s$^{-1}$. Similarly, for a true RV of $100$ \ms\ the bias would be less than $0.4$ \ms. These results are consistent across other absorption features considered. For additional discussion about applying the same analysis to other NSO absorption features, see Appendix \ref{appendix_modmisspec}.

\subsection{Nonparametric Template Estimation} \label{template_est}

Since the HGRV method models the difference in normalized flux, we need to have a template spectrum that approximates the quiet spectrum of a star with no stellar activity. In \changes{principal}, if one knows the approximate effective temperature, surface gravitational acceleration, metallicity, microturbulent velocity, and the elemental abundances of the star \changes{with high precision}, a synthetic spectrum could be produced at the proper resolution to give such a template \citep{sneden12}. However, \changes{in practice, these stellar parameters and the atomic line transition data are not well known enough to make this feasible. Therefore, we take a data-driven approach.}

\changes{The method we propose for estimating the template is to stack all normalized, barycentric corrected, observed spectra across time epochs and fit a smooth curve to the combined spectrum to estimate a representative spectrum.
The time sampling of the spectra can affect how well the estimated template approximates the true template.  For example, two of the possible extremes in the sampling are if all the observations are at the same orbital phase or if the observations are uniform across all phases.  
The estimated template under these extremes are not likely to affect the end result of the HGRV approach so this template estimation method is sufficient for our purposes.\footnote{Using the stacking and smoothing template estimate approach with time sampling that is approximately uniform across all phases of an exoplanet's orbit may lead to slightly broader features in the estimated template.  However, broadening tends to be primarily an even effect and so would not significantly hinder the RV estimation using the HGRV method, which fits an odd function ($\psi_{1}$) to the difference flux in Equation \eqref{bigXvariable}. Time sampling carried out in such a way that the observations occur at approximately the same phase of an exoplanet's orbit should not have this broadening of features. However, a constant RV offset may be present between the estimated template spectrum and all observed spectra. Because the same estimated template is used for each observation and only relative RV estimates are needed, this offset should not influence the fitted orbital parameters.
} 
} 

All observed spectra are stacked together, and we fit a local regression curve to this combined spectrum with a Gaussian kernel. We use local quadratic, instead of local linear, regression in order to better model the cores of absorption features. In practice we only fit at most $8$ \AA\ of the combined spectrum at a time, choosing an optimal bandwidth through generalized cross-validation for each section. This allows the computation to be parallelized. It also allows the bandwidth to be locally adaptive and take account of how absorption features are narrower on the blue end of the spectrum compared to the red end.  An advantage of this approach is that when stacking all observed spectra the wavelength solutions do not need to match across epochs, further minimizing the role of interpolation.
}

\changes{
\section{Simulation Studies}

This section includes two simulation studies based on the proposed methodology. The first is related to the template estimation approach, and the second compares properties of the RV estimation using the HGRV method with those of the commonly used CCF method.

\subsection{Template Estimation} \label{template_est_sim}
A nice feature of the HGRV approach is that no pre-specified template is required because the template spectrum is estimated from the full time-series of spectra using local quadratic regression (see Section \ref{template_est}). The estimated template contains both bias and variance, and we investigate the overall root mean squared error (RMS) through simulation. Furthermore, we consider how the RMS changes with the number of spectra and the S/N. Finally, we explore how the time-sampling cadence affects the estimated template.

For a star's true template with normalized flux $\tau$, and estimated template with normalized flux $\hat{\tau}$, we define the RMS as
\begin{equation}
\mathrm{RMS}\left(\hat{\tau}\right) \ =\ \sqrt{\dfrac{1}{n} \sum\limits_{i=1}^{n} (\tau_{i} - \hat{\tau}_{i})^{2}}. 
\end{equation}

For our simulation we use a version of the NSO spectrum that we smooth through local quadratic regression that approximately represents the quiet solar spectrum with infinite S/N. We also use cubic spline interpolation to give this smoothed NSO spectrum the same wavelength solution as the 51 Pegasi spectrum observed by EXPRES on Julian Day (JD) $2458641.952$. For a given number of observed spectra, $N$, each with a given S/N, our simulation consists of the following steps: (i) sample time epochs $t_{1}, . . . , t_{N}$ where $t_{k}\ \sim\ \mathrm{iid}\ \mathrm{Uniform}(0, 2\pi)$, (ii) calculate RV's $v_{r,1},..., v_{r,N} $ where  $v_{r,k} = 10 \mathrm{sin}(t_{k})$, (iii) simulate $N$ observed spectra with wavelength axis Doppler-shifted using Equation \eqref{dopplershift_classical} with RV $v_{r,k}$, and normalized flux axis with independent Poisson noise at the given S/N (where the noise is added to the un-normalized flux), (iv) apply the template estimation method described in Section \ref{template_est} and calculate the resulting RMS$\left(\hat{\tau}\right)$.

In our simulations, the number of spectra, $N$, ranges from $1$ to $31$ (in steps of $2$) and the S/N ranges from $100$ to $250$ (in steps of $10$). For each pair of values we perform the simulation $50$ independent times and calculate the average, and standard deviation, of the RMS. Each of these $50$ represents a different cadence. For computational purposes we do not use the entire spectrum for this simulation. Instead, we use the wavelength window $5240 - 5245$ \AA \ for our simulation. We also ran the same simulation on the wavelength window $(4965, 4970)$ which has a higher density of absorption features, as well as the window $(6381, 6386)$ which has a lower absorption feature density. The results for these additional windows are similar to the first window.  The results for the window $5240 - 5245$ \AA \ are summarized in Figure \ref{sim_tempest} which shows the average RMS$\left(\hat{\tau}\right)$ on the left panel, and the standard deviation of the RMS$\left(\hat{\tau}\right)$ on the right, for each pair of S/N and number of spectra.

\begin{figure}[h!]
\centering
\includegraphics[scale=0.75]{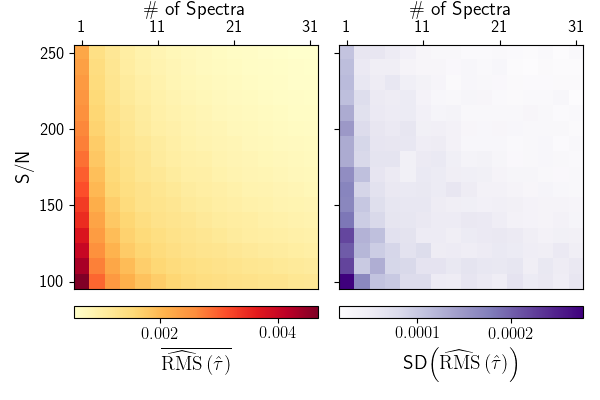}
\caption{Simulation study results for estimating the template spectrum between $5240$ and $5245$ \AA. For each S/N and number of spectra, $N$, 50 simulations were carried out each with a different cadence. Each simulation involved estimating the template with local quadratic regression and calculating the RMS. The left plot shows the average, and the right plot shows the standard deviation, of the RMS across the 50 simulations for each pair of S/N and $N$. The plots share the same vertical-axis.}
\label{sim_tempest}
\end{figure}

The left plot in Figure \ref{sim_tempest} illustrates that once the number of spectra reaches approximately $21$, the average RMS$\left(\hat{\tau}\right)$ of the estimated template is below approximately $0.001$ (which represents a S/N of about $1000$) for any S/N above $100$. On the other hand, if all observed spectra had a S/N above $200$ (which is often true of EXPRES spectra), one would only need about $11$ spectra to reach this template estimation precision. Furthermore, by examining the differences between the true template and individual instances of an estimated template, the residuals showed no obvious systematic bias within the wavelength bounds of absorption features. The same plot also shows that the RMS$\left(\hat{\tau}\right)$ is more affected by the number of spectra than the S/N in this example.

The right plot in Figure \ref{sim_tempest} illustrates how the RMS$\left(\hat{\tau}\right)$ varies due to the differing cadences in the 50 samples used for each pair of S/N and number of spectra. The simulation suggests that, as expected, the greatest differences are found when using only one spectrum. The variation is minimal for $11$ or more spectra and a S/N above $150$.

\subsection{RV Estimation} \label{rv_est_sim}
To investigate the accuracy of the HGRV method, especially at low velocities, we simulate spectra with a known RV and estimate the RMS of $\hat{v}_{r}$. By design, this simulation ignores astrophysical effects on RV-precision from stellar activity, analyzing the error contribution from modeling alone. To estimate this RMS, we use
\begin{equation}
    \widehat{\mathrm{RMS}}\left( \hat{v}_{r} \right) = \sqrt{\dfrac{1}{n}\sum\limits_{i=1}^{n} \left( \hat{v}_{r,i} - v_{r} \right)^{2}} \label{eqn:rms_hat}
\end{equation}
where $n$ is the number of simulations at RV $v_{r}$. The square of $\widehat{\mathrm{RMS}}\left( \hat{v}_{r} \right)$ can be decomposed into the sum of the variance and squared bias of $\hat{v}_{r}$ as well. To get a more detailed summary of our simulation we also estimate the standard deviation (SD) with
\begin{equation}
    \widehat{\mathrm{SD}}\left( \hat{v}_{r} \right) = \sqrt{\dfrac{1}{n}\sum\limits_{i=1}^{n} \left( \hat{v}_{r,i} - \bar{v}_{r} \right)^{2}} \label{eqn:sd_hat}
\end{equation}
where $\bar{v}_{r}$ is the average estimated velocity, and estimate the bias with 
\begin{equation}
    \widehat{\mathrm{Bias}}\left( \hat{v}_{r} \right) = \bar{v}_{r} - v_{r} \ . \label{eqn:bias_hat}
\end{equation}

We explore how the RMS$(\hat{v}_{r})$, Bias$(\hat{v}_{r})$, and SD$(\hat{v}_{r})$ vary with S/N and $v_{r}$. Our simulation takes $5$ equally spaced values of S/N $100, 150, ..., 300$ and $4$ values of $v_{r}$ equally spaced on a log scale from $0.01$ to $100\ m/s$. For each pair of S/N and  $v_{r}$ values, we use the estimated template spectrum for 51 Pegasi to simulate $2000$ independent spectra with the proper Doppler shift given by Equation \eqref{dopplershift_classical}. Each such simulation consists of using cubic splines to interpolate the shifted, oversampled, and high S/N template to the same wavelength solution as the observed 51 Pegasi spectrum from EXPRES on JD $2458639.958$ (see Section \ref{51peg_section} for more details) and including Poisson noise of the specified S/N. The results for obtaining each $\hat{v}_{r}$ with the HGRV method are shown in Figure \ref{hgrv_rms_sim}.

\begin{figure}[h!]
\centering
\includegraphics[scale=0.67]{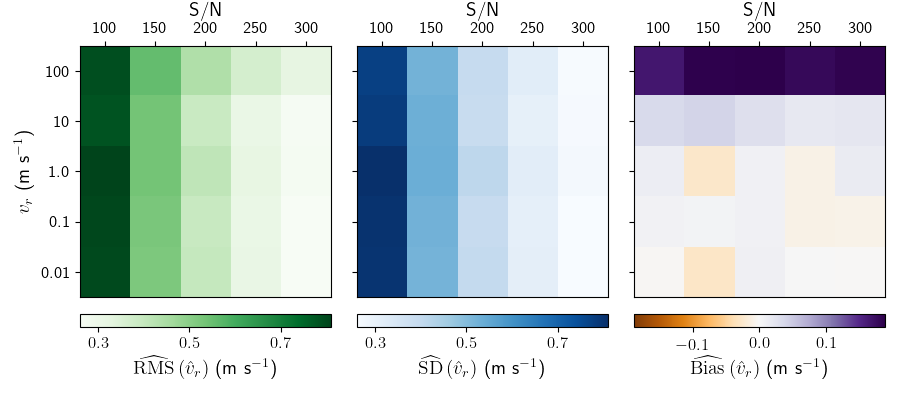}
\caption{The results for applying the HGRV method to spectra simulated from the estimated 51 Pegasi template spectrum. The left, middle, and right panels show the estimated RMS, SD, and bias of the estimated RV respectively. All three panels share the same vertical axis that represents the true RV each spectrum was simulated with. The S/N of the simulated spectra are given by the horizontal axis on top of each panel. The color scale for each panel is represented by the colorbar below it. Each pair of S/N and  $v_{r}$ involved $2000$ independent simulations to estimate the three quantities.}
\label{hgrv_rms_sim}
\end{figure}

The left panel of Figure \ref{hgrv_rms_sim} illustrates that the HGRV method is able to obtain a precision less than $0.3$ \ms\ when the S/N is approximately $250$ or higher, at least in the small RV regime. Additionally, the right panel of Figure \ref{hgrv_rms_sim} builds upon the model misspecification simulation done in Section \ref{model_misspecification} and informs us that combining many (non-Gaussian) absorption features in the HGRV method does not lead to an amplified systematic bias. We also find that the bias is somewhat proportional to the true RV. Furthermore, the SD contributes significantly more to the overall RMS than whatever bias may be present at the RV and S/N considered here.\footnote{We also performed the same simulation with a S/N of $1000$ and a RV of $1$ \ms\ (again using the estimated 51 Pegasi template spectrum and simulating $2000$ independent spectra). This simulation gave an estimated RMS of $0.077$ \ms, an estimated SD of $0.077$ \ms, and an estimated bias of $2.5 \times 10^{-3}$ \ms. This demonstrates that the HGRV method has the capability of obtaining a RV precision less than $0.1$ \ms.}

We also run the same simulation, estimating the RV with the CCF method as used in the EXPRES pipeline \citep{petersburg2020} with the HARPS G2 mask. Since the CCF method returns an absolute RV, rather than a relative RV, we first calculate the RV given for the estimated 51 Pegasi template with no noise ($-33168.5399$ \ms) and subtract this offset from all estimated RV's from the simulation. We then compare the estimated bias, SD, and RMS of the two methods at each pair of S/N and $v_{r}$. Figure \ref{hgrv_ccf_rmsdiff} shows the difference in RMS between the HGRV and CCF methods. Since every pair of S/N and $v_{r}$ in Figure \ref{hgrv_ccf_rmsdiff} shows a negative RMS difference, this suggests that the HGRV method has higher RV-precision than the CCF approach in this regime. %The EXPRES team finds that their FM code likewise gives a similarly improved fitting.

\begin{figure}[h!]
\centering
\includegraphics[scale=0.68]{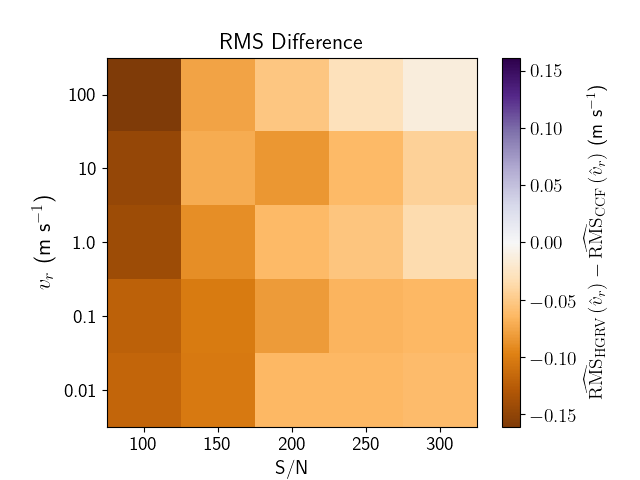}
\caption{The difference between the HGRV and CCF RMS for each pair of S/N and true $v_{r}$. Each pair consisted of $2000$ independent simulations for each method. The difference is indicated on the right by the color bar which is centered at $0.0$ \ms, and demonstrates the higher RV-precision of the HGRV method.}
\label{hgrv_ccf_rmsdiff}
\end{figure}

As a more detailed summary of the RMS improvement of the HGRV as demonstrated by Figure \ref{hgrv_ccf_rmsdiff}, the difference in the estimated SD and absolute bias (the sum of squares of which equal the squared RMS) is shown in Figure \ref{hgrv_ccf_varbiasdiff}.

\begin{figure}[h!]
\centering
\includegraphics[scale=0.68]{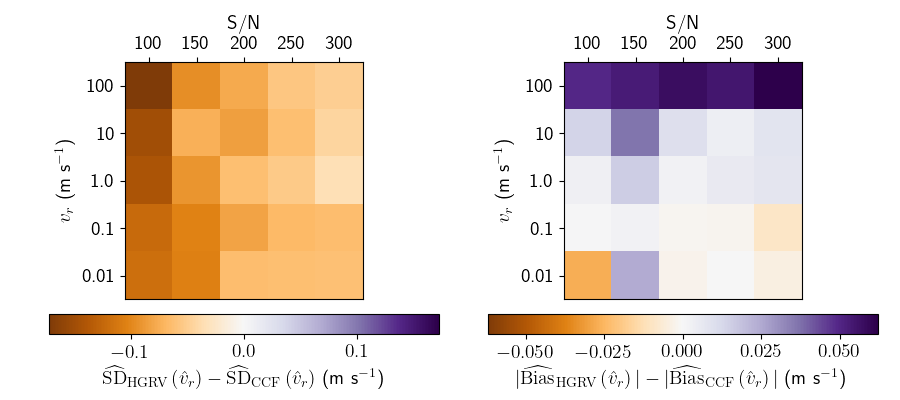}
\caption{The difference between the HGRV and CCF standard deviation and absolute bias for each pair of S/N and true $v_{r}$. Each pair consisted of $2000$ independent simulations for each method. The differences are indicated below each panel by the color bars which are centered at $0.0$ \ms. }
\label{hgrv_ccf_varbiasdiff}
\end{figure}

Figures \ref{hgrv_ccf_rmsdiff} and \ref{hgrv_ccf_varbiasdiff} inform us that the HGRV method is an example of the statistical phenomenon where a small increase in bias reduces the overall RMS. The greatest difference in RMS between the HGRV and CCF methods appears to be at low S/N.

To check the stability of this simulation, we used the wavelength solution for the 51 Pegasi spectrum from EXPRES observed on JD $2458804.588$ instead of the wavelength solution from JD $2458639.958$ used above. Running the HGRV and CCF approach each with $2000$ independent simulations with $v_{r} = 1$ \ms\ and a S/N of $200$ produced an RMS difference of $-0.094$ \ms. All estimated RVs from the CCF and HGRV methods for these simulations are provided in the repository \href{https://github.com/parkerholzer/hgrv_method}{$ https://github.com/parkerholzer/hgrv\underline{\ }method $}.

}

\section{Applications to 51 Pegasi data} \label{51peg_section}
51 Pegasi is the first \changes{main-sequence star similar to the Sun} discovered to possess an exoplanet \citep{mayor95}. The exoplanet has been found to have a RV semi-amplitude of $55.57 \pm 2.22$ \ms and orbital period of $4.2292 \pm 0.0003$ days \citep{mayor95, marcy97, wang11, bedell19}. To test the proposed HGRV method, we use data recently collected for 51 Pegasi by EXPRES \citep{jurgenson16, petersburg2020}. The recent spectrograph of EXPRES corrects for many of the instrumental effects that prior observations of 51 Pegasi were unable to avoid, allowing for greater  precision of derived RV. \changes{Our dataset consists of $56$ observed spectra from JD $2458639$ to $2458805$ (June 5, 2019 to Nov. 18, 2019). The S/N of these spectra ranges from $89$ to as high as $385$, but most are close to $200$ (see Table \ref{51pegrvs} for more details). These spectra have wavelength solutions that differ and do not consist of equally spaced pixels.}

\subsection{Data Corrections}
The raw data collected by the spectrograph do not have a flat continuum. This is in part due to the star's temperature causing more photons to be emitted at certain wavelengths than others. It is also due to instrumental effects such as the theoretical blaze function \citep{barker84, xu19}. \changes{To correct for these effects, we adopt the normalization from the EXPRES pipeline provided with each spectrum \citep{petersburg2020}.}

\changes{We also correct for the effects of the Earth's motion around the Sun by adopting the barycentric corrected wavelength solution provided with each observed spectrum by the EXPRES pipeline \citep{blackman17, blackman2020, petersburg2020}. Without the barycentric wavelengths provided by the EXPRES team, our derivation of RV would incur errors at the level of tens of cm s$^{-1}$.}

Finally, we correct for absorption features due to the Earth's atmosphere, often referred to as tellurics. Since the spectrograph is ground-based, the light from the star passes through the Earth's atmosphere, causing the presence of additional absorption features in the spectrum that are not representative of the target star. To correct for these tellurics, \changes{we use the model provided by EXPRES with each spectrum that was} created using the approach of \cite{leet19}. \changes{Although one could potentially divide out shallow tellurics to approximately correct for them with such a model, we take a more conservative approach and mask out all pixels with a telluric model normalized flux less than $1.0$.} %We also mask out pixels in the wavelength region of $5860 - 5915$ \AA \ as this region is known to have saturated, and therefore poorly modeled tellurics.

\begin{figure}[h!]
\centering
\includegraphics[scale=0.68]{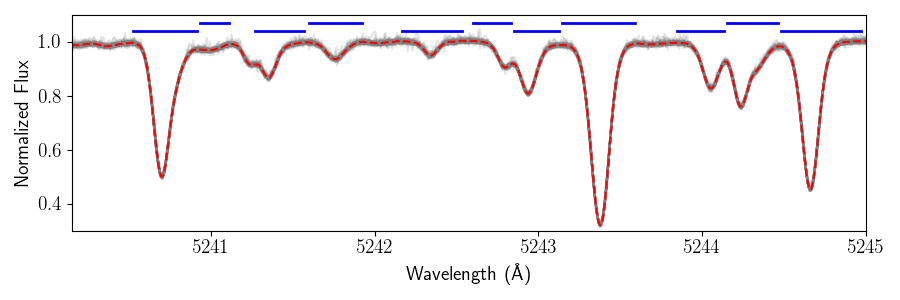}
\caption{A subset of the estimated template spectrum calculated from 51 Pegasi data is shown in the red dashed line on top of all observed spectra used in the calculation (shown in gray). The feature bounds that result from running Algorithm \ref{aff_algorithm} on the estimated template spectrum are also shown in blue horizontal lines. The full spectrum goes from $4470-6800$ \AA, but for visualization only $5240-5245$ \AA\  are displayed. \changes{The error bars of the estimated template  between $4850$ and $6800$ \AA\ (i.e., the wavelengths used in the analysis) have a median of $5.2 \times 10^{-4}$ and a $99$th percentile of $1.1 \times 10^{-3}$.}}
\label{51pegtemplatespec}
\end{figure}

Because a spectrum covers over $3000$ \AA\ of wavelength, the spectrograph collects the data in \changes{(partially overlapping)} wavelength orders stacked onto the rectangular detector. \changes{Therefore, we begin by stitching all orders of a given epoch together to create a single array of wavelength and normalized flux. To stitch two neighboring orders together in their overlapping region, we use cubic-spline interpolation to give the same wavelength solution to both orders in the overlap region \citep{meszaros13}. We then take the (point-wise) weighted average of the normalized flux in the overlap region of the two orders. Since the signal decreases at the edge of each order due to the instrumental blaze function, we set the weights for this averaging to decrease linearly for a given order as we get closer to the edge of the order. After applying this stitching to all neighboring orders we have a full observed spectrum for each epoch. 

We then proceed to estimate the template spectrum by way of local quadratic regression as described in Section \ref{template_est}. A small wavelength window of the estimated template spectrum that is calculated from the 51 Pegasi data is shown in Figure \ref{51pegtemplatespec}.}

Once we have the high S/N \changes{estimated} template spectrum we can use it in Algorithm \ref{aff_algorithm} to find absorption \changes{feature wavelength bounds}. The tuning parameters of the algorithm that were found through the \changes{optimization process described in Appendix \ref{appendix_aff}} were \changes{$m = 7$}, $\alpha = 0.05$, and $\eta = 0.07$ while eliminating any features with a line depth less than $0.015$. The algorithm finds a total of \changes{$4190$} features between wavelengths $4470$ \AA \ and \changes{$6800$} \AA. The results of this are also indicated in Figure \ref{51pegtemplatespec} for the section of the spectrum displayed. Note that when neighboring features are strongly blended together, Algorithm \ref{aff_algorithm} may either count both as a single feature or only pick out one of the two.

\subsection{Absorption Feature Parameters}

In order to use Equation \eqref{bigXvariable} and estimate the RV, we need to get estimates of the Gaussian parameters $d_{i}$, $\mu_{i}$, and $\sigma_{i}$ for each absorption feature $i$ using the high S/N \changes{estimated} template spectrum. To do so we use the Trust Region Reflective algorithm \citep{branch99}, which allows for initialization and bounds for each parameter to be fitted in non-linear least-squares. For absorption feature $i$ we initialize the Gaussian amplitude $d_{i}$ at one minus the minimum flux attained by the \changes{estimated template} spectrum within the wavelength bounds of feature $i$, the Gaussian center $\mu_{i}$ is initialized at the wavelength for which this minimum flux is attained, and the Gaussian spread $\sigma_{i}$ is initialized at one-fifth the width of the wavelength window for feature $i$. The bounds on the Gaussian amplitude are set to be $\left[0, 1\right]$, the Gaussian center is restricted to be within the wavelength bounds for feature $i$, and the Gaussian spread is lower-bounded by $0$ and upper-bounded by the width of the wavelength window for feature $i$.

\begin{figure}[h!]
\centering
\includegraphics[scale=0.68]{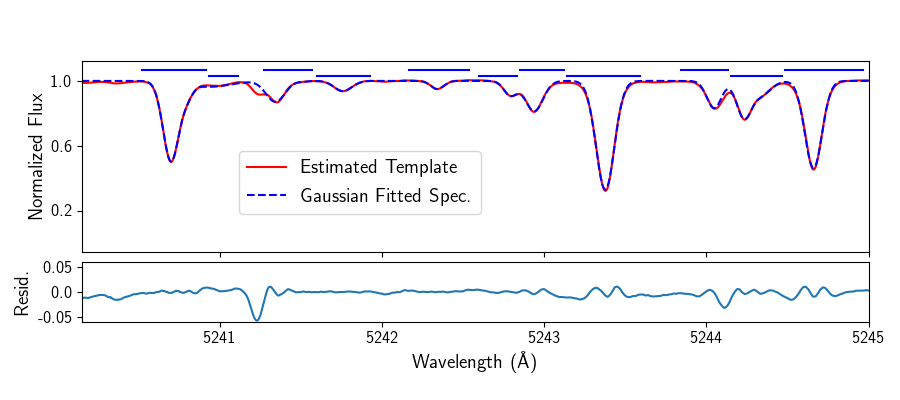}
\caption{The \changes{estimated template} spectrum for 51 Pegasi is shown in solid red with the spectrum that approximates it as a sum of Gaussians shown in dashed blue. The full spectra go from $4470-6800$ \AA, but for visualization only $5240-5245$ \AA\  are displayed. All absorption features in this wavelength range were well-fitted with Gaussians within the feature wavelength bounds. Portions of the spectrum that are poorly fitted with the sum of Gaussians are not contained within wavelength bounds of detected features, indicated with horizontal blue solid lines. The residual difference is shown below the main plot with the same Wavelength axis and a magnified vertical axis.}
\label{51peggaussfit}
\end{figure}

For computational purposes, we do not optimize the Gaussian parameters for all absorption features simultaneously. Instead, we estimate the parameters of one absorption feature by simultaneously optimizing that feature with its two neighboring features. If the resulting fit has a MSE within the wavelength bounds of the feature that is high\footnote{We consider a MSE to be high if it is greater than four multiples of the median MSE.}, which particularly happens when two strongly blended spectral lines are counted as one absorption feature, we try fitting a sum of two Gaussians to it. If this still does not give a good fit, we eliminate the respective feature \changes{so as to minimize the effects of model misspecification analyzed in Section \ref{model_misspecification}}. Out of the $4174$ absorption features detected by Algorithm \ref{aff_algorithm}, $3868$ were well-fitted with one or two Gaussians. An example of the fit model spectrum is shown in Figure \ref{51peggaussfit}. Most of the features that were eliminated at this stage were strongly blended with one or more neighboring features.

\subsection{Results}

To derive the RV for each epoch, we first limit the spectrum to the wavelength region $4850 - 6800$ \AA. \changes{While the wavelength solution is excellent from $5000$ to $7000$ \AA\ due to the laser frequency comb of EXPRES spanning that region \citep{blackman2020, petersburg2020}, and increasingly poor outside that window, we find that the spectra are acceptable for our purposes down to about $4850$ \AA.} Below $4850$ \AA \ the noise of the spectra increases and wavelengths above $6800$ \AA \ have too many strong telluric features. Limiting to this wavelength region reduces the number of absorption features from $3868$ to $2796$. We furthermore eliminate any pixels in the spectrum that are not contained in the wavelength windows of these $2796$ features.

\begin{figure}[h!]
\centering
\includegraphics[scale=0.75]{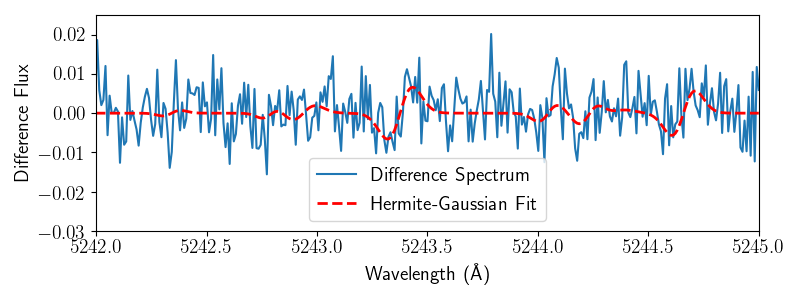}
\caption{The difference spectrum between the \changes{estimated template} and the spectrum observed on \changes{June 7, 2019 (JD $2458641.452$)} by EXPRES is shown in solid blue. The curve fitted according to Equation \eqref{bigXvariable} is shown in dashed red. For visualization, only $5242 - 5245$ \AA\ is shown.}
\label{diffspec_mod}
\end{figure}

After using cubic-splines to interpolate the \changes{high S/N, oversampled, estimated template} spectrum to the wavelength solution of the observed spectrum for a given epoch\footnote{This is the only time in the proposed method that interpolation takes place.} \citep{meszaros13}, we calculate the difference spectrum between the two. We then transform each wavelength $x_{i}$ using the sum, $\sum\limits_{j = 1}^{n} - \dfrac{\sqrt{\sqrt{\pi}} d_{j} \mu_{j}}{c \sqrt{2 \sigma_{j}}} \psi_{1}\left( x_{i} ; \mu_{j}, \sigma_{j} \right)$, from Equation \eqref{bigXvariable}. This transformation uses all fitted Gaussian parameters, after which we model the difference flux \changes{across the full stitched spectrum} as a function of this new variable using \changes{weighted} least-squares regression without an intercept to get the \changes{single} RV estimate, $\hat{v}_{r}$.\changes{\footnote{The usual regression diagnostics should be considered here (e.g., investigating extreme outliers or points with high leverage).  No issues were found in this application to 51 Pegasi.}} 
The standard error of $\hat{v}_{r}$ is also easily estimated by the usual least-squares approach. On average across the epochs, this standard error is approximately \changes{$0.52$ \ms}. An example of what the difference spectrum looks like in the interval $5242 - 5245$ \AA, together with the fitted Hermite-Gaussian model, is shown in Figure \ref{diffspec_mod}.

\changes{For our analysis we used the same 47 observations that were analyzed by the EXPRES team in \cite{petersburg2020} to estimate the orbital parameters.  Several available observations were excluded by the EXPRES team due to low S/N or failure of the laser frequency comb (see \citealt{petersburg2020} for details).  
The estimated RV's for all available 51 Pegasi EXPRES spectra using the proposed HGRV method are given in Table \ref{51pegrvs} of Appendix \ref{appendix_51peg}.
 Using the noted 47 EXPRES observations and the RV's estimated from the HGRV method, we compare the orbital parameters and the overall RV curve fit to those of the CCF method and the FM approach of \cite{petersburg2020}.
}

\changes{The exoplanet orbiting 51 Pegasi has been found to have an eccentricity that is nearly zero \citep{marcy97, wang11, bedell19, petersburg2020} implying an orbit that is nearly circular.} For a nearly circular planetary orbit, the host star's RV will behave approximately as a sine curve over time. Therefore, we use the Levenberg-Marquardt optimization algorithm \citep{more78} to fit a sine curve to the derived RV using 
\begin{eqnarray}
v_{r}(t) = K sin \left(\dfrac{2 \pi}{P} t + \phi \right) + b \ .
\label{rvcurve} 
\end{eqnarray}

\begin{figure}[h!]
\centering
\includegraphics[scale=0.8]{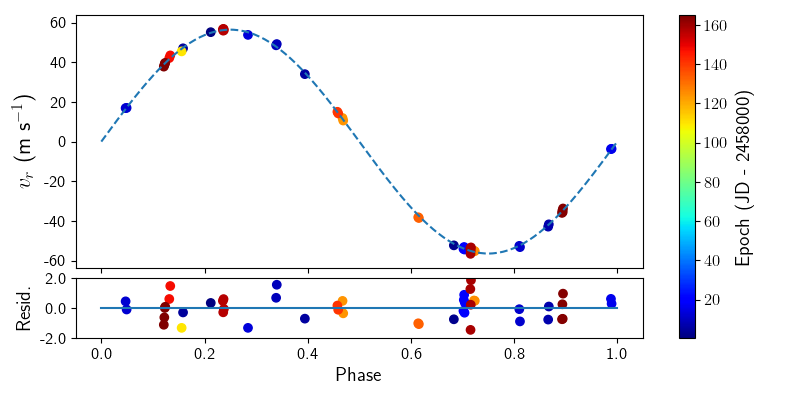}
\caption{The RV's derived for 51 Pegasi by the HGRV method, plotted as a function of \changes{orbital phase} with solid points \changes{whose color indicates the epoch according to the colorbar on the right}. All error bars are smaller than the size of the points. The fitted sine curve from Equation \eqref{rvcurve} is  also shown in a blue dashed curve using the HGRV values from Table \ref{rvfitpars}. The residuals are shown in the magnified window at the bottom and have the same units (\ms) as the plotted RV's.}
\label{RVfit}
\end{figure}

The semi-amplitude ($K$) is initialized at $55.5$ \ms\ and the period ($P$) at $4.23$ days. The phase ($\phi$), representing a horizontal shift of the sine curve, and the RV offset ($b$), giving the vertical shift, are both initialized at $0$. \changes{To account for instrumental changes to EXPRES, $b$ is allowed to be different before and after August, 2019.} The optimization converges to the fit parameters given in Table \ref{rvfitpars}\footnote{The fitted values of the two offsets are not given in Table \ref{rvfitpars} since they are expected to differ significantly between the three methods. The HGRV and FM methods give the RV relative to an estimated template, whereas the CCF method gives the RV relative to a pre-specified mask.}, and the results of this fitting are shown in Figure \ref{RVfit}. Therefore, the HGRV estimation method recovers the well-known parameters for 51 Pegasi. \changes{The only pair of parameters that had a significant correlation were the phase, $\hat{\phi}$, and the period, $\hat{P}$, which was $-0.813$. All other pairs had a correlation magnitudes less than $0.25$.}

\begin{table}[h!]
\centering
\begin{tabular}{|| c | c | c | c ||}
\hline
 & HGRV & CCF & FM \\ [0.5ex]
\hline
\hline
$\hat{K}$ & $56.48\ \pm \ 0.16$ \ms & $56.20\ \pm \ 0.19$ \ms & $56.17\ \pm \ 0.18$ \ms \\
\hline
$\hat{P}$ & $4.2308\ \pm \ 0.0001$ days & $4.2304\ \pm \ 0.0002$ days & $4.2306\ \pm \ 0.0002$\\
\hline
$\hat{\phi}$ & $-1.333\ \pm \ 0.006$ & $-1.326\ \pm \ 0.007$ & $-1.331\ \pm \ 0.007$\\
\hline
$RMS$ & $0.774$ \ms & $0.936$ \ms & $0.902$ \ms \\[.1ex]
\hline

\end{tabular}
\caption{Fit parameters of Equation \eqref{rvcurve} for 51 Pegasi.}
\label{rvfitpars}
\end{table}
 
\vspace{1cm}
Table \ref{rvfitpars} also gives the fit parameters from using the RV's \changes{estimated from the CCF and FM methods in \cite{petersburg2020} for the $47$ observations. Similar to the simulation study in Section \ref{rv_est_sim}, the reduced RMS demonstrates the ability of the HGRV method to outperform the traditional CCF approach.}

Including all $56$ available spectra gives an estimated $\hat{K} = 56.38 \pm 0.16$ \ms, $\hat{P} = 4.2308 \pm 0.0001$ days, $\hat{\phi} = -1.327 \pm 0.005$, and an RMS$= 0.858$ \ms.

\section{Discussion}

\changes{In this paper we introduce a new approach to estimate the RV in stellar spectra for exoplanet detection that we call the HGRV method. This method works by modeling the differences between observed normalized spectra and an estimated template spectrum. Even though this difference spectrum  visually appears to be nothing more than noise (e.g., see Figure \ref{diffspec_mod}), there is still an important Doppler signal present. By assuming that absorption features are \changes{approximately} Gaussian and that $v_{r}\ <\ 500$ \ms, the HGRV method is able to identify this small signal. The application to 51 Pegasi using spectra from EXPRES provides an example of how the HGRV-estimated RV's produce a lower RMS in the overall Keplerian fit than the classical CCF approach. Furthermore, the simulation study of Section \ref{rv_est_sim} demonstrates that at low RV, characteristic of Earth-like exoplanets orbiting Sun-like stars, the HGRV approach has higher RV-precision than the CCF.

Theorem \ref{maintheorem} implies that the difference flux, imposed on a Gaussian absorption feature by a planetary Doppler shift, can almost entirely be explained as a constant multiple of $\psi_{1}$. This reduces RV estimation to linear regression with no intercept, where the estimated coefficient is the estimated RV. Therefore, the RV can be interpreted as a proportionality constant between the difference flux and an explanatory variable expressed as a linear combination of first-degree generalized Hermite-Gaussian functions (see Equation \eqref{bigXvariable}). 

One of the benefits of the HGRV method is the simplification to linear regression, allowing for straight-forward statistical inference on the estimated RV.
Additionally, linear regression allows heteroskedasticity to be easily addressed with weighted least squares.

Interpolation is only used for stitching together the orders of each observed spectrum, and for getting the estimated template spectrum on the same wavelength solution as each observed spectrum. However, the interpolation for stitching orders can be fully avoided by taking each order out to the midpoint of the overlapping regions rather than using weighted averages. Alternatively, each order could be considered on its own as a way to fully avoid stitching orders. Furthermore, the template can be produced with the same wavelength solution as any observed spectrum by making these wavelengths the target in the local quadratic regression, therefore removing the need for later interpolation.

We also observed in the 51 Pegasi example that the HGRV method is relatively robust to inaccurate normalization. For example, the difference flux between the observation at JD $2458639.958$  and the estimated template has a visually identifiable offset from zero, but including this observation's estimated RV in the orbital parameter estimation of Equation \eqref{rvcurve} slightly reduced the model's RMS. This robustness may be due to how, on the scale of individual absorption features, inaccurate normalization is approximately an even effect. More work is needed, however, to confirm this general robustness.

An important feature of the HGRV method that also arises from its use of linear regression is its potential to be extended for disentangling Keplerian velocities due to exoplanets from photospheric velocities due to the star itself. The convective motion and magnetic activity of stars
lead to stellar activity in the form of starspots, granulation, faculae, etc.  which add red noise to the spectra of stars that can hide a true Doppler-shift or temporarily mimic a RV \citep{saar97, queloz01, desort07, meunier10}. Stellar activity can impose a false RV of approximate magnitude $1$ \ms\ for quiet stars \citep{hatzes02, lagrange10, isaacson10} to hundreds of \ms\ for the most active \citep{saar97, paulson04}. While efforts have been made to model this activity (e.g., \citealt{tuomi13, rajpaul15, delisle18}), as well as use alternative forms of the cross-correlation method to correct for activity (e.g., \citealt{queloz01, simola19}), these have had limited success in disentangling it from a true Doppler shift at RV's below $1$ \ms\ \citep{dumusque17}.

One way the HGRV method could potentially be utilized for disentangling stellar activity from Keplerian Doppler shifts is by approximately orthogonalizing these two effects. The general idea behind this is to find a way by which stellar activity affects absorption features and a Doppler shift does not. \cite{davis17} uses principal components analysis to show that, at least according to simplified models of the Sun,  the signals of stellar activity and a Doppler shift are distinguishable. Therefore, stellar activity would change a Gaussian absorption feature in a way that requires more Hermite-Gaussian terms than just $\psi_{1}$, whereas Theorem \ref{maintheorem} states that (at least at low RV) a Doppler shift would not. One could then use observations from either the Sun (e.g., \citealt{dumusque14}) or a star with high stellar activity levels (e.g., \citealt{giguere16}) to model $c_{1}$ in Equation \eqref{hermgauss_decomposition} as a function of the higher-degree coefficients, and remove the RV component that is due only to stellar activity. 
This is possible because the Hermite-Gaussian functions are orthogonal, and therefore as long as the blending between neighboring absorption features is small, a sum of higher-degree Hermite-Gaussian functions would be approximately orthogonal to the sum of first-degree Hermite-Gaussian functions. These ideas are the topic of future work.

}

The proposed method does have the limitation that at high values of RV, $c_{1}$ in Equation \eqref{hermgauss_decomposition} is no longer the only coefficient that is significantly non-zero (see Figure \ref{coef_vs_rv}), therefore, the HGRV method would not work well. Fortunately, very few exoplanets, none of which are Earth-like, exert such a large RV on their host star. But values of RV well above $500$ \ms\ easily arise when considering \changes{binary star systems}.

An improvement that could potentially be made to the proposed method is to relax the assumption of absorption features being Gaussian shaped. The advantage of using this assumption is that its derivative is a constant multiple of a basis function in the well known orthonormal Hermite-Gaussian basis. It is this orthogonality that potentially will allow us to orthogonalize the effects of stellar activity and a Doppler-shift. \changes{Furthermore, this assumption allows us to quantify with Theorem \ref{maintheorem} the approximation error of our model.} In order to replace the Gaussian assumption with a more general shape and potentially still model out stellar activity, one may need to have the derivative of the new shape be a basis function in another orthonormal basis.

Data and \changes{Python3} code associated with this work can be found at\\ \href{https://github.com/parkerholzer/hgrv_method}{$ https://github.com/parkerholzer/hgrv\underline{\ }method $}. \changes{The HGRV method is also implemented in the open source R package \textit{rvmethod}}.

\section{Conclusion}
By using the mathematical property that Doppler-shifting a Gaussian is nearly the same as adding a first-degree Hermite-Gaussian function, we propose a new method for estimating a Doppler shift in the spectrum of a star. Under the assumptions that the spectrum's absorption features can be well approximated by a sum of Gaussians and that the true RV is not too large in magnitude, the problem of estimating a RV in the spectrum can be \changes{simplified} to \changes{weighted} linear regression with no intercept. By testing this new method on recently collected, high-resolution spectra from EXPRES for the star 51 Pegasi we recover the well known orbital parameters \changes{with an overall RMS ($0.774$ \ms) below that of the traditional CCF method ($0.936$ \ms). This is only possible because the barycentric corrected wavelengths were provided by the EXPRES team. Furthermore, simulation studies demonstrate the ability of the HGRV method to outperform the CCF approach, giving an RV-prevision RMS that is up to approximately $15$ cm s$^{-1}$ lower than the CCF. This includes at the level of RV that is characteristic of Earth-like exoplanets orbiting Sun-like stars (i.e. $0.1$ \ms). } Unlike many other RV estimation algorithms, the HGRV method easily allows for statistical inference on the estimated RV, does not rely heavily on interpolation, takes account of the \changes{functional relationship} in neighboring pixels, and has a natural extension that could potentially be used to model out the effects of stellar activity.

\section*{Acknowledgements}
The authors gratefully acknowledge support through NSF-AST 1616086 and NASA XRP 80NSSC18K0443. \changes{This work used the EXtreme PREcision Spectrograph (EXPRES) that was designed and commissioned at Yale with financial support by the U.S. National Science Foundation under MRI-1429365 and ATI-1509436 (PI D. Fischer). The authors also gratefully acknowledge the EXPRES team for building this high fidelity instrument, providing the stellar spectra of 51 Pegasi and the benchmark radial velocities derived with their CCF and FM codes, and for helpful discussions. We also thank the Associate Editor and the two referees who provided valuable feedback and suggestions while reviewing this paper.  LLZ gratefully acknowledges support from the National Science Foundation Graduate Research Fellowship under Grant No. DGE1122492.  These results made use of the Lowell Discovery Telescope at Lowell Observatory. Lowell is a private, non-profit institution dedicated to astrophysical research and public appreciation of astronomy and operates the LDT in partnership with Boston University, the University of Maryland, the University of Toledo, Northern Arizona University and Yale University. We thank the Yale Center for Research Computing for guidance and use of the research computing infrastructure.
}

%\bibstyle{apalike}
\bibliography{HGRVNotes}
\bibliographystyle{apalike}
\nocite{*}

\appendix

\changes{
\section{Details of Absorption Feature Finder Algorithm } \label{appendix_aff}

Various algorithms already exist for detecting spectral features, particularly for emission lines in spectra of galaxies. However, they contain some limitations that make them unsuitable for the proposed methodology. For example, some were developed for absorption features of specific elemental species or line types\footnote{The central wavelength of each spectral line corresponds to a particular electron state transition of atoms responsible for absorbing photons in the stars photosphere. These central wavelengths depend on the species of the absorbing atom and its ionization state. }  \citep{frank08, zhao19}, require experimental supervision \citep{labutin13}, partially consist of extensive human intervention and physical insight \citep{sharpee03}, or assume the features are sparse and well-separated \citep{tonegawa15}. 

More importantly, these algorithms lack an important component needed for our analysis: estimating not just the central wavelength at which the feature occurs, but also the wavelength bounds that contain the feature. \cite{dumusque18} approaches this by taking a fixed number of pixels around each feature center, but acknowledges that these windows could be further optimized. The reason for this is because a fixed pixel count for each wavelength window does not take into account different sizes of absorption features nor blends between neighboring features. \cite{cretignier20} improves upon this by allowing the number of pixels to vary for each feature but, by restricting the windows to be symmetric about the minimum, does not account for effects of line blends. Our proposed algorithm improves upon this by using an approach that accounts for these blends.

Our proposed Algorithm \ref{aff_algorithm} works as follows. For a given pixel index $i$, let $\Lambda_{l,i}$ and $\Lambda_{r,i}$ be the wavelength regions of size $m$ pixels to the left and right of the wavelength for pixel $i$, $x_{i}$, respectively. Also, let $Y_{l,i}$ and $Y_{r,i}$ be the corresponding flux regions. Algorithm \ref{aff_algorithm} uses least-squares regression on each region to estimate coefficients $\beta_{0,l}$ and $\beta_{1,l}$ for the left region in addition to $\beta_{0,r}$ and $\beta_{1,r}$ for the right region (see Algorithm \ref{aff_algorithm} for the model). If $\beta_{1,l}$ is found to be negative and $\beta_{1,r}$ positive with statistical significance, then $x_{i}$ is considered a statistically significant minimum. At this point we apply a Bonferroni correction by using the significance level $\alpha/2$ for each slope. Algorithm \ref{aff_algorithm} then proceeds outwards in wavelength until the estimates are no longer statistically significant, at which point the central wavelength of the window is taken as a feature bound. To further avoid the drawbacks of multiple testing, we eliminate any detected absorption features that do not have a depth above a certain threshold. We note, however, that multiple testing is not a concern since our goal is to find absorption features, and we do not use the statistical significance beyond the detection of the features.

It was found that when $m$ is too small, many false absorption features are detected. When $m$ is too large, many small features are missed. Even though similar effects come from $\alpha$ and $\eta$ being too large or small, the effects appeared more sensitive to $m$. For fixed values of $\alpha$ and $\eta$, we adjusted $m$ until the number of detected features was maximized. At this point we increased or decreased $\alpha$ if many small features were missed or many false features were detected. If many blended features were detected as single features or the wavelength bounds did not encompass full absorption features, we decreased or increased $\eta$, respectively, and repeated the full process. 

When applying Algorithm \ref{aff_algorithm} to the NSO spectrum, we get the results shown in Figures \ref{aff_spectrum} and \ref{aff_missed}. 
Figure \ref{aff_spectrum} displays the portion of the spectrum that was not contained in any detected absorption features and compares it to the full spectrum. Figure \ref{aff_missed} displays some examples of absorption features that were missed by the algorithm.
These figures illustrate that $97.7\%$ of the squared deviation from $1.0$ in the normalized flux is accounted for by the $64.4\%$ of the spectrum contained in the wavelength bounds given by the algorithm. 

\begin{figure}[h!]
\centering
\includegraphics[scale=0.7]{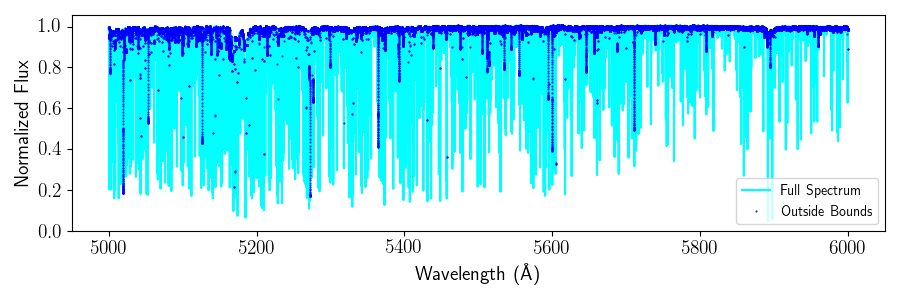}
\caption{The full NSO spectrum used in testing Algorithm \ref{aff_algorithm}. Normalized flux is plotted against the wavelength. The full spectrum is plotted in light blue. The thick dark blue points indicate the portions of the spectrum that are not contained in any of the wavelength bounds given by the algorithm.}
\label{aff_spectrum}
\end{figure}

It is also noticeable that some absorption features are missed by the algorithm, some of which are deep. Most of these were missed because, as illustrated in Figure \ref{aff_missed}, the features are strongly blended in a way that makes the slope in either direction at the core statistically insignificant. There are likely ways to improve upon this aspect of the algorithm, and we leave this to future work.

\begin{figure}[h!]
\centering
\includegraphics[scale=0.7]{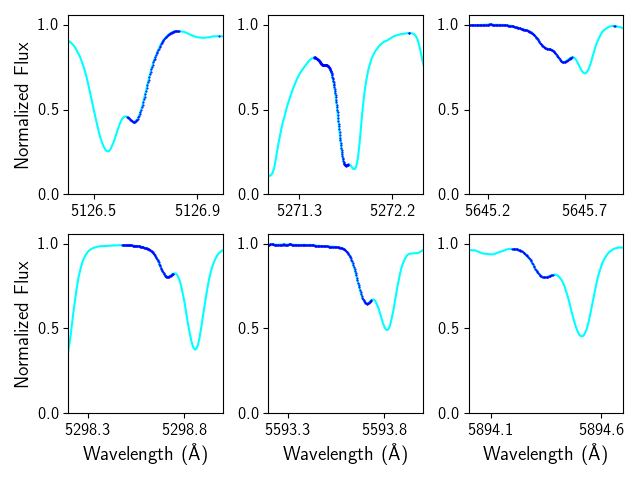}
\caption{Six of the absorption features in the NSO that were missed by Algorithm \ref{aff_algorithm}. Normalized flux is plotted against wavelength. The full spectrum is shown in light blue, and portions not included in any of the wavelength bounds given by the algorithm is shown in dark blue.}
\label{aff_missed}
\end{figure}

To analyze how the minimum line depth parameter depends on the S/N of the spectrum, we extend the false positive rate simulation done with a S/N of $500$ described in Section \ref{aff_section}. For each S/N from $250$ to $1500$ in equal steps of $250$, we take the NSO spectrum between $5000$ and $6000$ \AA\ and replace the flux axis with noise $20$ independent times. We then apply Algorithm \ref{aff_algorithm} to each of the $20$ resulting spectra with parameters $m\ =\ 25$, $\alpha\ =\ 0.01$, and $\eta\ =\ 0.05$. We then collect all detected absorption features from the $20$ spectra.

The total count of false absorption features detected ranged from $51$ to $56$ and showed no association with the S/N level. Furthermore, the depth of these false features is illustrated in Figure \ref{aff_fdr}. The recommended minimum line depth parameter, $0.015 \times \dfrac{500}{\mathrm{S/N}}$, is also shown.

\begin{figure}[h!]
\centering
\includegraphics[scale=0.7]{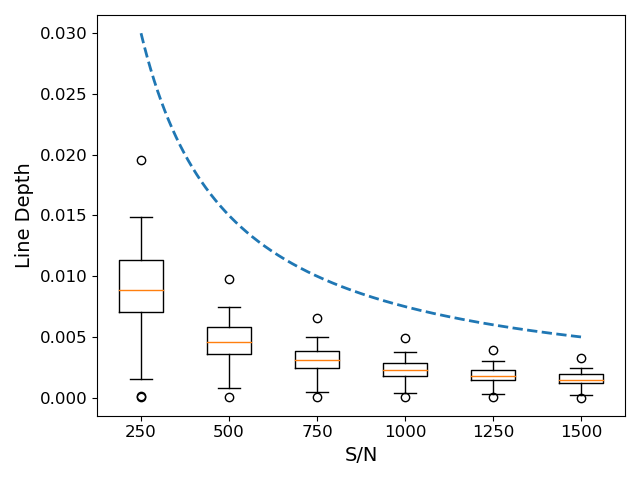}
\caption{Results from our simulation of the false positive rate of Algorithm \ref{aff_algorithm} at various S/N, shown on the horizontal axis. The distribution of line depths for these false positives is represented by box plots according to the vertical axis. The count of false positives remained approximately constant at $1$ absorption feature per $363$ \AA\ for each S/N. The dashed line represents our recommended value for the minimum line depth parameter in the algorithm given by the expression $0.015 \times \dfrac{500}{\mathrm{S/N}}$.}
\label{aff_fdr}
\end{figure}

}

\section{Proofs of Lemmas 1-4} \label{appendix_proofs}

\begin{proof}
(of Lemma \ref{intk_lemma})\\
Choose constants $a \in \mathbb{R}^{+}$, $b,c \in \mathbb{R}$. Then, using integration by parts, we have that
\begin{flalign}
&& I_{1}(a,b,c) &= %\int\limits_{-\infty}^{\infty} x e^{-\left(ax^{2} + bx + c\right)}dx %= \int\limits_{-\infty}^{\infty} x e^{-a\left(x + \dfrac{b}{2a} \right)^{2}}  e^{\left( \dfrac{b^{2}}{4a} - c \right)}dx && \text{by definition}\\
%&& &= e^{\left( \dfrac{b^{2}}{4a} - c \right)} \left[ \int\limits_{-\infty}^{\infty} \left( x + \dfrac{b}{2a} \right) e^{-a\left(x + \dfrac{b}{2a} \right)^{2}} dx - \dfrac{b}{2a} \int\limits_{-\infty}^{\infty} e^{-a\left(x + \dfrac{b}{2a} \right)^{2}} dx \right] && \\
%&& &
%= 
e^{\left( \dfrac{b^{2}}{4a} - c \right)} \left[ \int\limits_{-\infty}^{\infty} u e^{-a u^{2}} du - \dfrac{b}{2a} \int\limits_{-\infty}^{\infty} e^{-a u^{2}} du \right] %&& \text{substituting } u = x + \dfrac{b}{2a} \\
%&& &= e^{\left( \dfrac{b^{2}}{4a} - c \right)} \left[ 0 - \dfrac{b}{2a} \sqrt{\dfrac{\pi}{a}} \right] && \\&& &
= - \dfrac{\sqrt{\pi} b}{2 a^{3/2}} e^{\left( \dfrac{b^{2}}{4a} - c \right)} && \text{ . } \\
&& I_{0}(a,b,c) &= \int\limits_{-\infty}^{\infty} e^{-\left(ax^{2} + bx + c\right)}dx %= \int\limits_{-\infty}^{\infty} e^{-ax^{2}} e^{- \left(bx + c\right)}dx && \text{by definition}\\
%&& &
%= \lim\limits_{z \rightarrow \infty} \left( -\dfrac{1}{b}e^{-\left( a x^{2} + bx + c \right)} |_{-z}^{z} \right) - \dfrac{2a}{b} \int\limits_{-\infty}^{\infty} x e^{-\left(ax^{2} + bx + c\right)}dx &&  \text{using integration by parts} \\
%&& &
= \dfrac{2a}{b} I_{1}(a,b,c) = \sqrt{\dfrac{\pi}{a}} e^{\left( \dfrac{b^{2}}{4a} - c \right)} && \text{ .} 
\end{flalign}
Now choose any $k \in \{ n \in \mathbb{N} : n \geq 2 \}$.
\begin{flalign}
&& I_{k-1}(a,b,c) &= %\int\limits_{-\infty}^{\infty} x^{k-1} e^{-\left( a x^{2} + bx + c \right)}dx = 
 \int\limits_{-\infty}^{\infty} x^{k-1} e^{-a x^{2}} e^{-(bx + c)}dx && \\%\text{by definition} \\
&& &= \lim\limits_{z \rightarrow \infty} \left[ -\dfrac{1}{b} x^{k-1} e^{-\left( a x^{2} + bx + c \right)} |_{-z}^{z} \right] + \dfrac{1}{b}\int\limits_{-\infty}^{\infty} \left( (k-1) x^{k-2} - 2ax^{k} \right) e^{-\left( a x^{2} + bx + c \right)}dx && \\
&& &= \dfrac{k-1}{b}I_{k-2}(a,b,c) - \dfrac{2a}{b}I_{k}(a,b,c) && \text{ . }
\end{flalign}
So we have that 
\begin{equation}
I_{k}(a,b,c) = -\dfrac{b}{2a} I_{k-1}(a,b,c) + \dfrac{k-1}{2a} I_{k-2}(a,b,c).
\end{equation}
\end{proof}

\begin{proof}
(of Lemma \ref{ck_lemma})\\
Since $g(x; \gamma) = \sum\limits_{n=0}^{\infty} c_{n}(\gamma) \psi_{n}(x ; \mu, \sigma)$ and $\psi_{n}(x ; \mu, \sigma)$ are orthonormal, we have that
\begin{eqnarray}
%\int\limits_{-\infty}^{\infty} \psi_{k}(x ; \mu, \sigma)g(x; \gamma) dx = \int\limits_{-\infty}^{\infty} \sum\limits_{n=0}^{\infty} c_{n}(\gamma)\psi_{k}(x ; \mu, \sigma) \psi_{n}(x ; \mu, \sigma) dx \\
%= \sum\limits_{n=0}^{\infty} c_{n}(\gamma) \int\limits_{-\infty}^{\infty} \psi_{k}(x ; \mu, \sigma) \psi_{n}(x ; \mu, \sigma) dx = \sum\limits_{n=0}^{\infty} c_{n}(\gamma) \mathds{1} \{k=n\} = c_{k}(\gamma) \ . 
c_{k}(\gamma) = \int\limits_{-\infty}^{\infty} \psi_{k}(x ; \mu, \sigma)g(x; \gamma) dx \ .
\label{coef_equation}
\end{eqnarray}

Choose any $k \in \{ n \in \mathbb{N} : n \geq 1 \}$. By using Equation \eqref{hermitepoly} for the $k$'th Hermite polynomial, we have that for $\varepsilon = \gamma - 1$,
\begin{flalign}
&& c_{k}(\varepsilon) &= \int\limits_{-\infty}^{\infty} g(x; \varepsilon) \psi_{k}(x ; \mu, \sigma)dx &&  \\
%&& &= \int\limits_{-\infty}^{\infty} \sqrt{\sigma \sqrt{\pi}}\left( \psi_{0}(x ; \mu, \sigma) - \psi_{0}((1+ \varepsilon) x ; \mu, \sigma)\right)\psi_{k}(x ; \mu, \sigma) dx && \\
&& &= \sqrt{\sigma \sqrt{\pi}} \int\limits_{-\infty}^{\infty} \psi_{0}(x ; \mu, \sigma)\psi_{k}(x ; \mu, \sigma)dx - \int\limits_{-\infty}^{\infty} e^{- \dfrac{1}{2\sigma^{2}} (x - \mu + \varepsilon x)^{2}} \psi_{k}(x ; \mu, \sigma)dx && \label{intermediate_eqn} \\
&& &= 0 - \int\limits_{-\infty}^{\infty} \dfrac{1}{\sqrt{\sigma 2^{k} k! \sqrt{\pi}}} H_{k}\left( \dfrac{x - \mu}{\sigma} \right) e^{-\dfrac{1}{2\sigma^{2}} \left[ (x -\mu + \varepsilon x)^{2} + (x - \mu)^{2} \right] }dx && \\
&& &= - \dfrac{\sqrt{\sigma}}{\sqrt{2^{k} k! \sqrt{\pi}}} \int\limits_{-\infty}^{\infty} H_{k}(u) e^{-\dfrac{1}{2} \left[ \left( u + \varepsilon \left( u + \dfrac{\mu}{\sigma} \right) \right)^{2} + u^{2} \right]}du && \\%\text{substituting } u = \dfrac{x-\mu}{\sigma} \\
&& &= - \dfrac{\sqrt{\sigma}}{\sqrt{2^{k} k! \sqrt{\pi}}} \int\limits_{-\infty}^{\infty} k! \sum\limits_{m=0}^{\lfloor k/2 \rfloor} \dfrac{(-1)^{m}}{m! (k-2m)!} (2u)^{k-2m} e^{-\dfrac{1}{2} \left[ \left( 2 + 2\varepsilon + \varepsilon^{2} \right) u^{2} + 2 \varepsilon \dfrac{\mu}{\sigma}(1 + \varepsilon) u + \varepsilon^{2}\dfrac{\mu^{2}}{\sigma^{2}} \right]}du &&  \\
&& &= - \dfrac{\sqrt{\sigma k! 2^{k}}}{\sqrt{\sqrt{\pi}}} \sum\limits_{m=0}^{\lfloor k/2 \rfloor} \dfrac{(-1)^{m}}{m! (k-2m)!} \dfrac{1}{4^{m}} \int\limits_{-\infty}^{\infty} u^{k-2m} e^{-\dfrac{1}{2} \left[ \left( 2 + 2\varepsilon + \varepsilon^{2} \right) u^{2} + 2 \varepsilon \dfrac{\mu}{\sigma}(1 + \varepsilon) u + \varepsilon^{2}\dfrac{\mu^{2}}{\sigma^{2}} \right]}du &&  \\
&& &= - \sqrt{\dfrac{\sigma k! 2^{k}}{\sqrt{\pi}}} \sum\limits_{m=0}^{\left \lfloor k/2 \right \rfloor} \dfrac{(-1)^{m}}{4^{m} m! (k-2m)!}I_{k-2m} \left( 1 + \varepsilon + \dfrac{\varepsilon^{2}}{2} , \dfrac{\varepsilon \mu}{\sigma}(1+\varepsilon), \dfrac{1}{2}\left( \dfrac{\varepsilon \mu}{\sigma} \right)^{2} \right) 
\end{flalign}

For $k=0$, the only difference is that the first integral in Equation \eqref{intermediate_eqn} becomes $1$ instead of vanishing. Therefore,

\begin{flalign}
&& c_{0}(\varepsilon) %&= \int\limits_{-\infty}^{\infty} g(x; \varepsilon) \psi_{0}(x ; \mu, \sigma)dx && \\
%&& &= \int\limits_{-\infty}^{\infty} \sqrt{\sigma \sqrt{\pi}}\left( \psi_{0}(x ; \mu, \sigma) - \psi_{0}((1+ \varepsilon) x ; \mu, \sigma)\right)\psi_{0}(x ; \mu, \sigma) dx && \\
%&& &= \sqrt{\sigma \sqrt{\pi}} \int\limits_{-\infty}^{\infty} \psi_{0}^{2}(x ; \mu, \sigma) dx - \sqrt{\sigma \sqrt{\pi}} \int\limits_{-\infty}^{\infty} \psi_{0}((1+ \varepsilon) x ; \mu, \sigma)\psi_{0}(x ; \mu, \sigma) dx && \\
%&& &= \sqrt{\sigma \sqrt{\pi}} - \dfrac{1}{\sqrt{\sigma \sqrt{\pi}}} \int\limits_{-\infty}^{\infty} e^{-\dfrac{1}{2\sigma^{2}}\left( (x-\mu)^{2} + (x - \mu +\varepsilon x)^{2} \right)} dx && \\
%&& &= \sqrt{\sigma \sqrt{\pi}} - \dfrac{1}{\sqrt{\sigma \sqrt{\pi}}} \int\limits_{-\infty}^{\infty} e^{-\dfrac{1}{2\sigma^{2}}\left( 2x^{2} -4x\mu + 2\mu^{2} + 2\varepsilon x^{2} -2 \varepsilon \mu x + \varepsilon^{2} x^{2} \right)} dx && \\
%&& &= \sqrt{\sigma \sqrt{\pi}} - \dfrac{1}{\sqrt{\sigma \sqrt{\pi}}} \int\limits_{-\infty}^{\infty} e^{-\dfrac{1}{\sigma^{2}}\left( \left( 1 + \varepsilon + \dfrac{\varepsilon^{2}}{2} \right) x^{2} - \left( 2\mu + \varepsilon \mu \right) x + \mu^{2} \right)} dx && \\
&= \sqrt{\sigma \sqrt{\pi}} - \dfrac{1}{\sqrt{\sigma \sqrt{\pi}}} I_{0}\left( \dfrac{1 + \varepsilon + \dfrac{\varepsilon^{2}}{2}}{\sigma^{2}}, -\dfrac{2\mu + \varepsilon \mu}{\sigma^{2}}, \left( \dfrac{\mu}{\sigma} \right)^{2} \right) %&& \\
%&& &= \sqrt{\sigma \sqrt{\pi}} - \dfrac{1}{\sqrt{\sigma \sqrt{\pi}}} \sqrt{\dfrac{\pi}{\dfrac{1 + \varepsilon + \varepsilon^{2}/2}{\sigma^{2}}}} e^{\left( \dfrac{\dfrac{(2 + \varepsilon)^{2}\mu^{2}}{\sigma^{4}}}{4\dfrac{1 + \varepsilon + \varepsilon^{2}/2}{\sigma^{2}}} - \dfrac{\mu^{2}}{\sigma^{2}}\right)} && \\
%&& &= \sqrt{\sigma \sqrt{\pi}} \left( 1 - \dfrac{1}{\sqrt{1 + \varepsilon + \dfrac{\varepsilon^{2}}{2}}} e^{\left(\dfrac{(2 + \varepsilon)^{2} \mu^{2}}{4 \sigma^{2}\left( 1+\varepsilon+\varepsilon^{2}/2 \right)} - \dfrac{\mu^{2}}{\sigma^{2}}\right)}\right) && \\
\end{flalign}

\end{proof}

\begin{proof}
(of Lemma \ref{ratio_lemma}) \\
Decompose as $g(x; \gamma) = \sum\limits_{n=0}^{\infty} c_{n}(\gamma) \psi_{n}(x ; \mu, \sigma)$.\\
Then 
\begin{equation}
\int_{-\infty}^{\infty} \left( g(x; \gamma) - c_{1}(\gamma) \psi_{1}(x ; \mu, \sigma)\right)^{2} dx
\end{equation}
\begin{align}
 &= \int_{-\infty}^{\infty} (g(x; \gamma))^{2} dx - 2c_{1}(\gamma) \int_{-\infty}^{\infty} g(x; \gamma) \psi_{1}(x ; \mu, \sigma)dx + c_{1}^{2}(\gamma)\int_{-\infty}^{\infty} \psi_{1}^{2}(x ; \mu, \sigma)dx \\ &= \int_{-\infty}^{\infty} (g(x; \gamma))^{2} dx - 2c_{1}^{2}(\gamma) + c_{1}^{2}(\gamma) = \int_{-\infty}^{\infty} (g(x; \gamma))^{2} dx - c_{1}^{2}(\gamma)
\end{align}
%So $\dfrac{\int_{-\infty}^{\infty} \left( g(x; \gamma) - c_{1}(\gamma) \psi_{1}(x ; \mu, \sigma)\right)^{2} dx}{\int_{-\infty}^{\infty} \left( g(x; \gamma) \right)^{2} dx} = 1 - \dfrac{c_{1}^{2}(\gamma)}{\int_{-\infty}^{\infty} \left( g(x; \gamma) \right)^{2} dx}$.
\end{proof}

\begin{proof}
(of Lemma \ref{final_lemma}) \\
From Lemmas \ref{intk_lemma} and \ref{ck_lemma} we have that, with $\varepsilon = \gamma -1$,
\begin{eqnarray}
c_{1}^{2}(\varepsilon) %= \dfrac{4\sqrt{\pi}\mu^{2}}{\sigma} \dfrac{\varepsilon^{2}(1+\varepsilon)^{2}}{\left( 2 + 2\varepsilon + \varepsilon^{2} \right)^{3}}e^{-\left( \dfrac{\mu}{\sigma} \right)^{2} \dfrac{\varepsilon^{2}}{2 + 2\varepsilon + \varepsilon^{2}}} 
= \varepsilon^{2}(1 + \varepsilon)^{2} h(\varepsilon) %\label{c1squared_eqn}
\end{eqnarray}
where
\begin{eqnarray}
h(\varepsilon) := \dfrac{4\sqrt{\pi}\mu^{2}}{\sigma} \dfrac{1}{\left( 2 + 2\varepsilon + \varepsilon^{2} \right)^{3}}e^{-\left( \dfrac{\mu}{\sigma} \right)^{2} \dfrac{\varepsilon^{2}}{2 + 2\varepsilon + \varepsilon^{2}}}\ .
\end{eqnarray}
We also have that
\begin{eqnarray}
\dfrac{\partial}{\partial \varepsilon} c_{1}^{2}(\varepsilon) = \left(4 \varepsilon^{3} + 6 \varepsilon^{2} + 2\varepsilon\right) h(\varepsilon) + \varepsilon^{2}(1 + \varepsilon)^{2} \dfrac{\partial h(\varepsilon)}{\partial \varepsilon}
\end{eqnarray}
and
\begin{eqnarray}
\dfrac{\partial^{2}}{\partial \varepsilon^{2}} c_{1}^{2}(\varepsilon) = \left(12 \varepsilon^{2} + 12\varepsilon + 2 \right) h(\varepsilon) + 2\left(4 \varepsilon^{3} + 6 \varepsilon^{2} + 2\varepsilon\right) \dfrac{\partial h}{\partial \varepsilon} + \varepsilon^{2}(1 + \varepsilon)^{2} \dfrac{\partial^{2} h}{\partial \varepsilon^{2}}\ .
\end{eqnarray}
Since $h(\varepsilon)$, $\dfrac{\partial h}{\partial \varepsilon}$, and $\dfrac{\partial^{2} h}{\partial \varepsilon^{2}}$ are all continuous at $0$, we have that 
\begin{eqnarray}
\lim\limits_{\varepsilon \rightarrow 0} c_{1}^{2}(\varepsilon) = 0, \\
\lim\limits_{\varepsilon \rightarrow 0} \dfrac{\partial}{\partial \varepsilon} c_{1}^{2}(\varepsilon) = 0, \\ 
\text{and } \lim\limits_{\varepsilon \rightarrow 0} \dfrac{\partial^{2}}{\partial \varepsilon^{2}} c_{1}^{2}(\varepsilon) = 2 \lim\limits_{\varepsilon \rightarrow 0} h(\varepsilon) = \dfrac{\sqrt{\pi}\mu^{2}}{\sigma} .
\end{eqnarray}
With $g(x; \mu, \sigma)$ as in Lemma \ref{ck_lemma}, we have that 
\begin{align}
\int\limits_{-\infty}^{\infty} g^{2}(x; \gamma) dx &= \int\limits_{-\infty}^{\infty} e^{-\dfrac{(x-\mu)^{2}}{\sigma^{2}}} dx + \int\limits_{-\infty}^{\infty} e^{-\dfrac{(\gamma x-\mu)^{2}}{\sigma^{2}}} dx - 2 \int\limits_{-\infty}^{\infty} e^{-\dfrac{1}{2\sigma^{2}} \left( \left( 1 + \gamma^{2} \right) x^{2} - 2\mu (1 + \gamma) x + 2\mu^{2}\right)} dx  \\
&= \sigma \sqrt{\pi} + \dfrac{\sigma}{\gamma}\sqrt{\pi} - 2 e^{-\dfrac{\mu^{2}}{\sigma^{2}}\left( 1 - \dfrac{(1+\gamma)^{2}}{2\left( 1 + \gamma^{2} \right)}\right)} \int\limits_{-\infty}^{\infty} e^{-\dfrac{1+\gamma^{2}}{2\sigma^{2}} \left( x - \dfrac{\mu(1+\gamma)}{1+\gamma^{2}}\right)^{2}} dx  \\
&= \sigma\sqrt{\pi}\left( 1 + \dfrac{1}{\gamma} - \dfrac{2^{3/2}}{\sqrt{1 + \gamma^{2}}} e^{-\dfrac{\mu^{2}}{\sigma^{2}}\left( 1 - \dfrac{(1+\gamma)^{2}}{2\left( 1 + \gamma^{2}\right)}\right) }\right)  \\
&= \sigma\sqrt{\pi}\left( 1 + \dfrac{1}{1+\varepsilon} - \dfrac{2^{3/2}}{\sqrt{2 + 2\varepsilon + \varepsilon^{2}}} e^{-\dfrac{\mu^{2}}{2\sigma^{2}} \dfrac{\varepsilon^{2}}{2 + 2\varepsilon + \varepsilon^{2}} }\right) \label{denom_eqn}
\end{align}
Therefore, $\lim\limits_{\varepsilon \rightarrow 0} \int\limits_{-\infty}^{\infty} g^{2}(x; \varepsilon) dx = 0$.
Furthermore, we have that 
\begin{multline}
\dfrac{\partial}{\partial \varepsilon} \int\limits_{-\infty}^{\infty} g^{2}(x; \varepsilon) dx = \sigma\sqrt{\pi}\left[ -\dfrac{1}{(1+\varepsilon)^{2}} + 2^{3/2} \left( \left(2 + 2\varepsilon + \varepsilon^{2} \right)^{-3/2}(1+ \varepsilon) \right. \right. \\ \left. \left. + \dfrac{\mu^{2}}{\sigma^{2}}\left(2 + 2\varepsilon+\varepsilon^{2}\right)^{-5/2} \left(2\varepsilon + \varepsilon^{2}\right)\right) e^{-\dfrac{\mu^{2}}{2\sigma^{2}} \dfrac{\varepsilon^{2}}{2+2\varepsilon+\varepsilon^{2}} } \right] \ . \label{1stderiv_gsquared}
\end{multline}
Hence, 
\begin{equation}
\lim\limits_{\varepsilon \rightarrow 0} \dfrac{\partial}{\partial \varepsilon} \int\limits_{-\infty}^{\infty} g^{2}(x; \varepsilon) dx = 0.
\end{equation}

Defining 
\begin{equation}
h(\varepsilon) := 2^{3/2} \left( \left(2 + 2\varepsilon + \varepsilon^{2} \right)^{-3/2}(1+ \varepsilon) + \dfrac{\mu^{2}}{\sigma^{2}}\left(2 + 2\varepsilon+\varepsilon^{2}\right)^{-5/2} \left(2\varepsilon + \varepsilon^{2}\right)\right),
\end{equation}
we have that $h(\varepsilon)$ is continuous and differentiable at $0$.

Therefore, since 
\begin{equation}
\lim\limits_{\varepsilon \rightarrow 0} e^{-\dfrac{\mu^{2}}{2\sigma^{2}} \dfrac{\varepsilon^{2}}{2+2\varepsilon+\varepsilon^{2}} } = 1
\end{equation}
and 
\begin{equation}
\lim\limits_{\varepsilon \rightarrow 0} \dfrac{\partial}{\partial \varepsilon} e^{-\dfrac{\mu^{2}}{2\sigma^{2}} \dfrac{\varepsilon^{2}}{2+2\varepsilon+\varepsilon^{2}} } = 0,
\end{equation}
we have that 
\begin{equation}
\lim\limits_{\varepsilon \rightarrow 0} \dfrac{\partial}{\partial \varepsilon} \left( h(\varepsilon) e^{-\dfrac{\mu^{2}}{2\sigma^{2}} \dfrac{\varepsilon^{2}}{2+2\varepsilon+\varepsilon^{2}} } \right) = \lim\limits_{\varepsilon \rightarrow 0 } \dfrac{\partial h(\varepsilon)}{\partial \varepsilon}.
\end{equation}

Since 
\begin{equation}
\lim\limits_{\varepsilon \rightarrow 0} \dfrac{\partial}{\partial \varepsilon} \left( \left(2 + 2\varepsilon + \varepsilon^{2} \right)^{-3/2}(1+ \varepsilon) \right) = -3 \cdot 2^{-5/2} + 2^{-3/2}
\end{equation}
and 
\begin{equation}
\lim\limits_{\varepsilon \rightarrow 0} \dfrac{\partial}{\partial \varepsilon} \left( \left(2 + 2\varepsilon+\varepsilon^{2}\right)^{-5/2} \left(2\varepsilon + \varepsilon^{2}\right)\right) = 2^{-3/2},
\end{equation}
we have that 
\begin{equation}
\lim\limits_{\varepsilon \rightarrow 0 } \dfrac{\partial h(\varepsilon)}{\partial \varepsilon} = \dfrac{\mu^{2}}{\sigma^{2}} - \dfrac{1}{2}.
\end{equation}

And since 
\begin{equation}
\lim\limits_{\varepsilon \rightarrow 0} \dfrac{\partial}{\partial \varepsilon} \left( \dfrac{-1}{(1 + \varepsilon)^{2}} \right) = 2, 
\end{equation}
we have from Equation \eqref{1stderiv_gsquared} that

\begin{align}
\lim\limits_{\varepsilon \rightarrow 0} \dfrac{\partial^{2}}{\partial \varepsilon^{2}} \int\limits_{-\infty}^{\infty} g^{2}(x; \varepsilon) dx &= \dfrac{3\sigma\sqrt{\pi}}{2} + \dfrac{\sqrt{\pi}\mu^{2}}{\sigma}\ .
\end{align}
So 
\begin{equation}
\lim\limits_{\varepsilon \rightarrow 0} \dfrac{c_{1}^{2}(\varepsilon)}{\int\limits_{-\infty}^{\infty} g^{2}(x; \varepsilon) dx} = \dfrac{\dfrac{\sqrt{\pi}\mu^{2}}{\sigma}}{\dfrac{\sqrt{\pi}\mu^{2}}{\sigma} + \dfrac{3\sigma\sqrt{\pi}}{2}} = \dfrac{1}{1 + \dfrac{3 \sigma^{2}}{2 \mu^{2}}}.
\end{equation}
\end{proof}

\changes{

\section{Model Misspecification Details} \label{appendix_modmisspec}

Following the same procedure as in Section \ref{model_misspecification}, we considered $100$ additional absorption features to analyze the effect of misspecifying their profile as Gaussian, five of which are displayed below in Figure \ref{modelmisspec_extra}. A Gaussian density shape is fit to each absorption feature, which is then Doppler-shifted by $50$ equally spaced values of RV from $1$ to $100$ \ms.  The RV is then estimated using the HGRV method. Most, but not all, of the additional features we analyzed lead to a slight overestimate of the RV. But for all $100$ of these additional features, the difference for a $1$ \ms\ RV is less than $1$ cm s$^{-1}$ away from the truth. Furthermore, the simulations in Section \ref{rv_est_sim} indicate that when combining the lines in the HGRV method, the overall bias is not greater than with individual lines.

\begin{figure}[h!]
\centering
\includegraphics[scale=0.5]{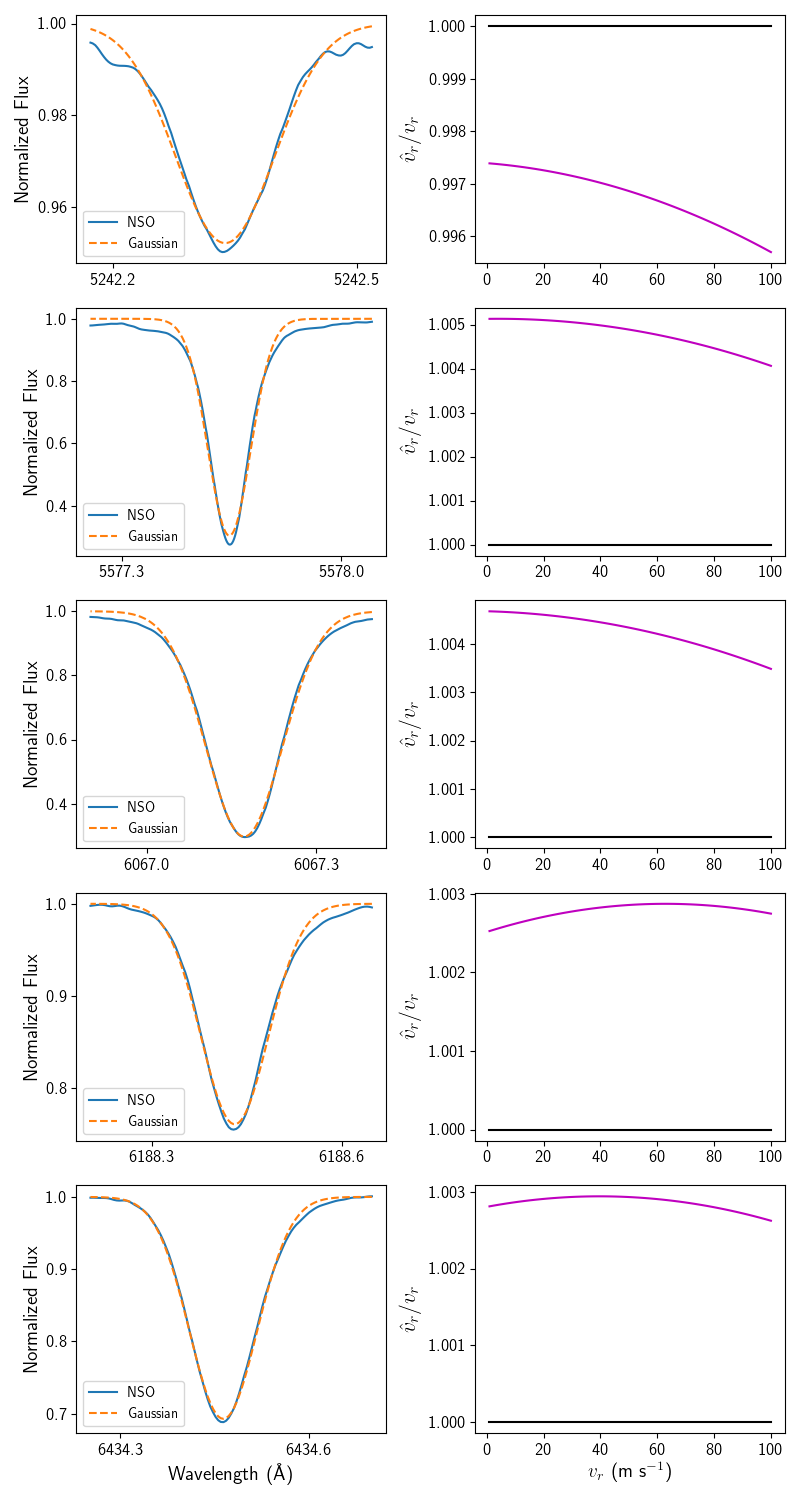}
\caption{Results for analyzing the effects of misspecifying the model of five different absorption features in the NSO spectrum as a Gaussian. The left panels show the feature in solid blue and the best fitted Gaussian in dashed orange. The right panels show the ratio of the RV estimated with Equation \eqref{bigXvariable} $\hat{v}_{r}$ (with $n=1$) and the true RV $v_{r}$.}
\label{modelmisspec_extra}
\end{figure}

\newpage

\section{51 Pegasi Radial Velocities}\label{appendix_51peg}

Here we give the RVs derived using the HGRV method on the $56$ observed spectra from EXPRES. 

\begin{longtable}{|| c | c | c ||}
%\centering

\hline \multicolumn{1}{|c|}{\textbf{MJD (days)}} & \multicolumn{1}{c|}{\textbf{RV (\ms)}} & \multicolumn{1}{c|}{\textbf{S/N}}  \\ \hline 
\endfirsthead

\multicolumn{3}{c}%
{{\bfseries \tablename\ \thetable{} -- continued from previous page}} \\
\hline \multicolumn{1}{|c|}{\textbf{MJD (days)}} & \multicolumn{1}{c|}{\textbf{RV (\ms)}} & \multicolumn{1}{c|}{\textbf{S/N}} \\ \hline 
\endhead

\hline \multicolumn{3}{|r|}{{Continued on next page}} \\ \hline
\endfoot

%\hline \hline
\endlastfoot
%\begin{tabular}
%\hline
% & RV (\ms) &  & In \cite{petersburg2020} \\ [0.5ex]
%\hline
%\hline
$58639.458442$ & $54.708\ \pm \ 0.404 $ & $385$\\
\hline
$58641.451749$ & $-52.850\ \pm \ 0.516$ & $179$ \\
\hline
$58641.457773^{\ast}$ & $-53.662\ \pm \ 0.710$ & $140$ \\
\hline
$58643.462180$ & $46.574\ \pm \ 0.521$ & $225$ \\ 
\hline
$58644.460959$ & $33.536\ \pm \ 0.512$ & $233$\\
\hline
$58646.455970$ & $-43.411\ \pm \ 0.444$ & $203$ \\
\hline
$58646.461286$ & $-42.241\ \pm \ 0.438$ & $204$\\
\hline
$58648.456163$ & $48.082\ \pm \ 0.505$ & $244$  \\
\hline
$58648.461529$ & $48.711\ \pm \ 0.498$ & $256$  \\
\hline
$58650.450235$ & $-53.092\ \pm \ 0.474$ & $199$  \\
\hline
$58650.455542$ & $-53.741\ \pm \ 0.486$ & $193$  \\
\hline
$58651.443961^{\ast}$ & $14.317\ \pm \ 1.130$ & $99$  \\
\hline
$58651.452932$ & $16.403\ \pm \ 0.431$ & $284$  \\
\hline
$58651.461117$ & $16.515\ \pm \ 0.519$ & $202$  \\
\hline
$58652.456394$ & $53.336\ \pm \ 0.653$ & $172$  \\
\hline
$58652.461797^{\ddagger}$ & $52.696\ \pm \ 0.667$ & NA  \\
\hline
$58655.432426$ & $-4.268\ \pm \ 0.459$ & $220$  \\
\hline
$58655.437704$ & $-4.148\ \pm \ 0.453$ & $222$  \\
\hline
$58657.456051^{\ast}$ & $11.484\ \pm \ 0.706$ & $142$  \\
\hline
$58657.461248^{\ast}$ & $11.792\ \pm \ 0.623$ & $157$  \\
\hline
$58658.453711$ & $-54.614\ \pm \ 0.361$ & $230$  \\
\hline
$58658.456675$ & $-53.959\ \pm \ 0.343$ & $247$  \\
\hline
$58658.459600$ & $-53.690\ \pm \ 0.341$ & $243$  \\
\hline
$58658.462634$ & $-54.958\ \pm \ 0.330$ & $257$ \\
\hline
$58658.465250$ & $-54.357\ \pm \ 0.350$ & $236$ \\
\hline
$58664.447934^{\ast}$ & $36.335\ \pm \ 1.171$ & $94$  \\
\hline
$58664.458268^{\ast}$ & $35.866\ \pm \ 1.270$ & $89$  \\
\hline
$58665.461782^{\dagger}$ & $43.919\ \pm \ 0.538$ & $214$  \\
\hline
$58749.221866$ & $47.001\ \pm \ 0.684$ & $163$  \\
\hline
$58749.227235^{\ddagger}$ & $48.181\ \pm \ 0.688$ & NA  \\
\hline
$58763.233618$ & $13.399\ \pm \ 0.466$ & $230$  \\
\hline
$58763.239194$ & $12.112\ \pm \ 0.455$ & $238$  \\
\hline
$58764.311548$ & $-53.651\ \pm \ 0.401$ & $244$  \\
\hline
$58764.318051$ & $-53.765\ \pm \ 0.372$ & $273$  \\
\hline
$58772.315903$ & $-36.680\ \pm \ 0.414$ & $234$  \\
\hline
$58772.321086$ & $-37.061\ \pm \ 0.420$ & $231$  \\
\hline
$58780.114819$ & $16.529\ \pm \ 0.464$ & $237$  \\
\hline
$58780.121270$ & $15.728\ \pm \ 0.462$ & $238$  \\
\hline
$58787.198050$ & $43.659\ \pm \ 0.557$ & $194$  \\
\hline
$58787.206110$ & $44.981\ \pm \ 0.504$ & $226$  \\
\hline
$58796.099263$ & $58.195\ \pm \ 0.546$ & $235$ \\
\hline
$58796.102083$ & $57.456\ \pm \ 0.544$ & $236$  \\
\hline
$58796.104824$ & $58.356\ \pm \ 0.546$ & $235$ \\
\hline
$58796.107532$ & $57.717\ \pm \ 0.543$ & $235$  \\
\hline
$58798.128178$ & $-52.396\ \pm \ 0.412$ & $234$  \\
\hline
$58798.129893$ & $-55.148\ \pm \ 0.411$ & $233$  \\
\hline
$58798.131622$ & $-53.502\ \pm \ 0.411$ & $231$  \\
\hline
$58798.133471$ & $-51.899\ \pm \ 0.409$ & $232$ \\
\hline
$58803.110815$ & $-34.492\ \pm \ 0.418$ & $233$  \\
\hline
$58803.114000$ & $-33.286\ \pm \ 0.418$ & $233$  \\
\hline
$58803.116558$ & $-34.086\ \pm \ 0.416$ & $233$  \\
\hline
$58803.118928$ & $-32.252\ \pm \ 0.413$ & $234$  \\
\hline
$58804.076698$ & $39.312\ \pm \ 0.503$ & $239$  \\
\hline
$58804.080907$ & $40.058\ \pm \ 0.502$ & $239$  \\
\hline
$58804.084687$ & $40.916\ \pm \ 0.504$ & $240$  \\
\hline
$58804.088298$ & $41.212\ \pm \ 0.503$ & $239$  \\%[1ex]
\hline \hline

%\vspace{1cm}

%\end{tabular}
\caption{Radial velocities derived from the HGRV method for 51 Pegasi. The first column gives the Modified Julian Day (MJD) which can be converted to JD by adding $2400000.5$ days. The second column gives the estimated RV with its standard error, and the third column identifies the S/N. In the first column, $\ast$ indicates that it was not included in \cite{petersburg2020} due to a S/N below $160$. $\dagger$ indicates that it was not included because the laser-frequency comb of the EXPRES spectrograph failed. $\ddagger$ indicates it wasn't included due to a charge transfer inefficiency in the spectrograph detector. A machine-readable version of this table is available on the online repository for this paper.
}
\label{51pegrvs}
\end{longtable}

The EXPRES spectra used to obtain these estimated RV's with the HGRV method came with the barycentric corrected wavelength solutions provided which we used. All 56 were used in estimating the template spectrum for 51 Pegasi, and used the same set of identified absorption features and Gaussian fits to this estimating template. 

}

\end{document}